\documentclass[twocolumn,aps,amssymb,footinbib]{revtex4-1}
\usepackage{lpic}
\usepackage{amssymb}
\usepackage{graphicx}
\usepackage{amsmath}
\usepackage{times}
\usepackage{subfigure}
\usepackage{feynmp}

\newcommand{\xx}{\mathbf{x}}

\newcommand{\qq}{\mathbf{q}}
\newcommand{\QQ}{\mathbf{Q}}
\newcommand{\kk}{\mathbf{k}}
\newcommand{\pp}{\mathbf{p}}

\newcommand{\GG}{\mathcal{G}}

\begin{document}
\begin{fmffile}{collective-polar-excitons-feyn}
\fmfset{thin}{0.75pt}
\fmfset{thick}{1.50pt}
\fmfset{arrow_len}{2.5mm}
\fmfset{dot_size}{1.5thick}
\fmfset{zigzag_len}{1.5mm}
\fmfset{zigzag_width}{0.75mm}
\fmfset{dash_len}{1mm}

\title{Collective phenomena in quasi-two-dimensional fermionic polar molecules:\\band renormalization and excitons}

\author{Mehrtash Babadi$^1$, Eugene Demler$^1$}
\affiliation{
$^1$ Physics Department, Harvard University, Cambridge, Massachusetts 02138, USA
}

\begin{abstract}
We theoretically analyze a quasi-two-dimensional system of fermionic polar molecules in a harmonic transverse confining potential. The renormalized energy bands are calculated by solving the Hartree-Fock equation numerically for various trap and dipolar interaction strengths. The inter-subband excitations of the system are studied in the conserving time-dependent Hartree-Fock (TDHF) approximation from the perspective of lattice modulation spectroscopy experiments. We find that the excitation spectrum consists of both inter-subband particle-hole excitation continuums and anti-bound excitons, arising from the anisotropic nature of dipolar interactions. The excitonic modes capture the majority of the spectral weight. We also evaluate the inter-subband transition rates in order to investigate the nature of the excitonic modes and find that they are anti-bound states formed from particle-hole excitations arising from several subbands. Our results indicate that the excitonic effects are present for interaction strengths and temperatures accessible in current experiments with polar molecules.
\end{abstract}

\maketitle

\section{Introduction}
In the past decade, much of the experimental and theoretical progress in the field of ultracold atoms~\cite{Lewenstein2007,Bloch2008,Ketterle2008} are motivated by the prospect of realizing novel strongly correlated many-body states of matter. In particular, experimental realization of quantum degenerate gases of fermionic polar molecules has witnessed a very rapid progress. By association of atoms via a Feshbach resonance to form deeply bound ultracold molecules~\cite{Lang2008,Deiglmayr2008}, a nearly degenerate gas of KRb polar molecules in their rotational and vibrational ground state has been recently realized~\cite{Ospelkaus2008,Ni2008,Ni2009,Ospelkaus2010,Ni2010}. The molecules can be polarized by applying a d.c. electric field, resulting in strong dipole-dipole inter-molecular interactions.

A strikingly new feature of systems of fermionic polar molecules is the anisotropy of dipolar interactions, making them unparalleled among the traditional condensed matter systems. Thus, experiments with polar molecules go beyond quantum simulation of effective theories motivated by electronic systems and aim at exploring a genuinely new domain of many-body quantum behavior, unique to dipolar interactions. Dipolar interactions can be utilized to generate long-range interactions of arbitrary shape using microwave fields~\cite{Buchler2007}, simulate exotic spin Hamiltonians~\cite{Micheli2006,Ortner2009} and are theoretically predicted to give rise to numerous interesting collective phenomena such as roton softening~\cite{Santos2003,Ronen2007,Wang2008}, supersolidity~\cite{Sengupta2005,Boninsegni2005,Capogrosso2009,Pollet2009,Burnell2009}, p-wave superfluidity~\cite{Bruun2008}, emergence of artificial photons~\cite{Tewari2006}, bilayer quantum phase transitions~\cite{Wang2007}, multi-layer self-assembled chains~\cite{Wang2006} for bosonic molecules, dimerization and inter-layer pairing~\cite{Potter2010,Baranov2011}, spontaneous inter-layer coherence~\cite{Lutchyn2009}, itinerant ferroelectricity~\cite{Lin2009}, anisotropic Fermi liquid theory and anisotropic sound modes~\cite{Sogo2009,Miyakawa2008,Chan2010,Ronen2010}, fractional quantum Hall effect~\cite{Baranov2005}, Wigner crystallization~\cite{Baranov2008}, density-wave and striped order~\cite{Yamaguchi2010,Sun2010}, biaxial nematic phase~\cite{Fregoso2009}, topological superfluidity~\cite{Cooper2009} and $\mathrm{Z}_2$ topological phase~\cite{Sun2009}, just to mention a few.

Despite the theoretical prediction of numerous exotic quantum many-body phenomena in polar molecules, realization and observation of many of these novel predictions are still an experimental challenge. At the time this paper was written, the coldest gas of fermionic polar molecules has been realized with KRb molecules at a temperature of $1.4\,T_F$~\cite{Ospelkaus2008,Ni2008,Ni2009,Ospelkaus2010,Ni2010}, where $T_F$ is the Fermi temperature. The majority of the mentioned quantum phenomena require strong suppression of thermal fluctuations, i.e. strong degeneracy condition ($T \ll T_F$).

A major obstacle towards further evaporative cooling of a large class of bi-alkali polar molecules (KRb, $\mathrm{LiNa}$, $\mathrm{LiK}$, $\mathrm{LiRb}$, and $\mathrm{LiCs}$) is the existence of an energetically allowed two-body chemical reaction channel~\cite{Zuchowski2010}, resulting in significant molecule losses in two-body scatterings. In a low temperature gas composed of a single hyperfine state, Fermi statistics blocks scatterings in the s-wave channel and the majority of scatterings take place through the p-wave channel. In unstructured three-dimensional clouds, the attractive head-to-tail dipolar interactions soften the p-wave centrifugal barrier and increase the cross section of reactive collisions. However, the rate of chemical reactions can be effectively suppressed by loading the gas into a one-dimensional optical lattice (or trap) and aligning the dipoles perpendicular to the formed {\em pancakes}. In such geometries, the incidence of head-to-tail scatterings is effectively suppressed due to the transverse confinement of the gas on one hand, and reinforcement of the p-wave barrier due to repulsive side-by-side dipolar interactions on the other hand~\cite{Ospelkaus2010,Quemener2010,Quemener2011}. Therefore, the preferred geometry to study reactive polar molecules is in tightly confined two-dimensional layers.

Recently, it has been shown that the suppression of chemical reactions of reactive fermionic molecules in confined geometries remains effective even in the quasi-two-dimensional limit, i.e. where not only the lowest, but also the first few excited subbands (transverse modes) are populated~\cite{Quemener2011}. On the other hand, occupation of higher subbands does not impose any difficulty on experiments with non-reactive molecules such as $\mathrm{NaK}$, $\mathrm{NaRb}$, $\mathrm{NaCs}$, $\mathrm{KCs}$, and $\mathrm{RbCs}$~\cite{Zuchowski2010}. The possibility of going beyond the single-subband limit opens a new window towards experimental and theoretical exploration of many-body physics of quasi-two-dimensional fermions with anisotropic interactions.

Besides being highly anisotropic, fermionic polar molecules and electronic systems differ in another important way: in contrast to the Coulomb interactions, dipole-dipole interactions dominate the kinetic energy in the high density limit. In $d$ dimensions, dipolar interaction scales as $n^{3/d}$ ($n$ is the density) while the kinetic energy scales as $n^{2/d}$, implying that their ratio scales as $n^{1/d}$, i.e. the interactions are more pronounced at higher densities. We define the following dimensionless quantity as a measure of dipolar interaction in two dimensions:
\begin{equation}
r_d \equiv \frac{m\,D^2\,n^{1/2}}{\hbar^2},
\end{equation}
where $D$ is the electric dipole moment of a single molecule.\\

The main goal of this paper is studying the inter-subband collective modes of quasi-two-dimensional polar molecules and to propose experimental signatures of such phenomena. At the time this paper was written, the dipolar interaction strengths accessible in the experiments belong to the weakly interacting regime (e.g. $r_d \approx 0.05$ in experiments of the group at JILA) and therefore, we restrict our analysis to the same regime. For small $r_d$, the normal liquid phase is expected to be the stable phase of the system.

In the first part of this paper, we study the effect of dipolar interactions on the single-particle energy dispersions by solving the Hartree-Fock equation. In order to simplify our analysis, we focus on a single pancake and neglect complications such as inter-layer tunneling and attractive inter-layer interactions in this study. This approximation is relevant to a well-separated stack of pancakes as well. The normal phase and the collective modes of fermionic polar molecules in the strictly two-dimensional case (single-subband limit) has been recently studied by different authors~\cite{Bruun2008,Ronen2010,Yamaguchi2010}. Therefore, we do not discuss the intra-subband collective modes here and instead, focus on genuinely quasi-two-dimensional phenomena.

\begin{figure}[h]
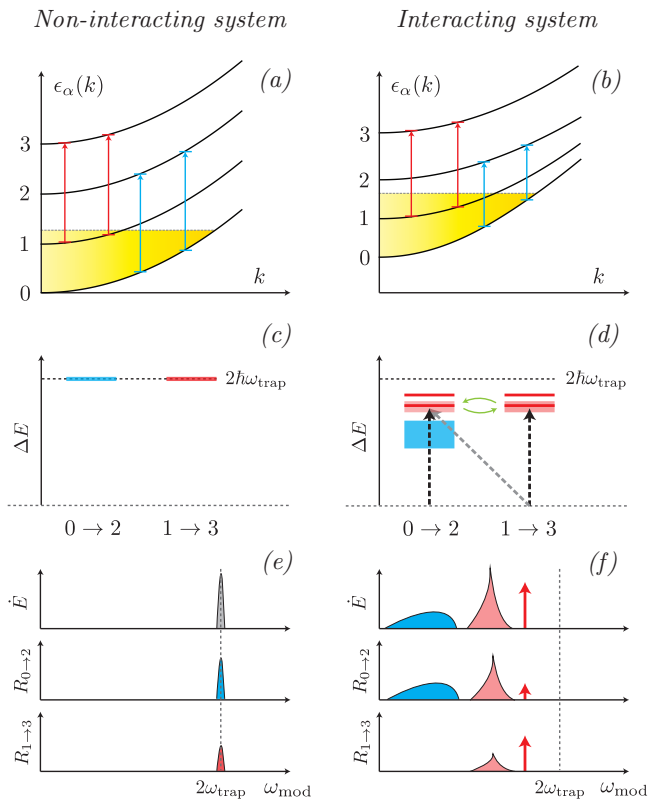

\begin{lpic}{figs/part1-intuition(8.5cm,)}
\end{lpic}
\caption{A comparison between quasi-two-dimensional fermions in the absence (left column) and presence (right column) of interactions. (a) and (b) show a schematic plot of the band structure. The blue and red arrows indicate p-h excitations from 0 to 2, and 1 to 3 subbands respectively, at different regions in the Fermi sea. (c) and (d) show a schematic representation of excited states probed in modulation spectroscopy experiments. The labels, $0 \rightarrow 2$ and $1 \rightarrow 3$, indicate the inter-subband transition associated to the excitation. The blue and red intervals denote p-h continuums and the red lines denote excitonic modes. (e) and (f) show the energy absorption rate, $\dot{E}$, and inter-subband transition rates, $R_{0\rightarrow 2}$ and $R_{1 \rightarrow 3}$, as a function of modulation frequency, $\omega_{\mathrm{mod}}$. The blue and red shaded continuums correspond to $0 \rightarrow 2$ and $1 \rightarrow 3$ inter-subband transitions respectively. The red spike denotes an exciton. The kink in the middle of the red shaded continuum is indicating of an exciton lying inside that continuum (see plots c and d). Refer to the text for more details.}
\label{fig:intuition}
\end{figure}

In the second part of this paper, we explore the inter-subband excitation spectrum from the lattice modulation spectroscopy perspective, an experimental technique originally developed for studying strongly correlated phases of cold atomic systems~\cite{Stoeferle2004,Iucci2006,Kollath2006,Sensarma2009,Tokuno2011}. We predict the energy absorption and inter-subband transition rates for an exponentially switched-on a.c. lattice modulation pulse in the conserving time-dependent Hartree-Fock (TDHF) approximation~\cite{BaymKadanoff1961}. We show that the excitation spectrum consists of both inter-subband particle-hole (p-h) excitation continuums and anti-bound excitons, arising from the anisotropic nature of dipolar interactions. We show that these many-body effects persist for weak interactions and temperatures of the order of the Fermi temperature, making their observation feasible in current experiments. Finally, we study the inter-subband particle transition rates in order to shed light on the nature of excitons. We find that in contrast to p-h excitation continuums which are associated to certain inter-subband transitions, the excitons are composite anti-bound states formed from p-h excitations arising from several subbands.\\

Before we embark on the formal development of the outlined program, we find it worthwhile to summarize the obtained results in a schematic way. Fig.~\ref{fig:intuition} shows a side by side comparison between non-interacting and interacting quasi-two-dimensional fermions with dipolar interactions. A typical plot of the band structure is shown in Fig.~\ref{fig:intuition}a and~\ref{fig:intuition}b. In the absence of interactions, however, the single-particle energy dispersions are quadratic and have a constant energy separation of $\hbar\omega_{\mathrm{trap}}$, where $\omega_{\mathrm{trap}}$ is the trap frequency (we neglect the anharmonicity of the trap in order to simplify the analysis). In the presence of interactions, the single-particle energy dispersions are renormalized and no longer remain quadratic.

We will see later that modulation of the optical lattice stimulates excitation of particles from populated subbands in equilibrium to their second next subband ($0 \rightarrow 2$, $1 \rightarrow 3$, etc). The excitation energy of such processes can be directly measured in the experiments. In the absence of interactions, all of the elementary inter-subband excitations have a constant energy cost of $2\hbar\omega_{\mathrm{trap}}$. The interactions modify the degenerate excited states dramatically, transforming the p-h excitation energies from their constant value of $2\hbar\omega_{\mathrm{trap}}$ to a collection of continuums and discrete collective modes (excitons), as shown in Fig.~\ref{fig:intuition}c and~\ref{fig:intuition}d. Also, the excitonic modes capture the majority of the spectral weight, leaving behind a small trace of p-h continuums.

The mechanism of energy absorption in lattice modulation experiments can be explored further by studying the inter-subband transition rates. If the modulation frequency lies within a certain inter-subband p-h excitation continuum, we expect to observe transitions only between the two involved subbands, leaving the population of other subbands unaffected. For example, the blue shaded continuum in the energy absorption spectrum in Fig.~\ref{fig:intuition}d and~\ref{fig:intuition}f is associated to $0 \rightarrow 2$ excitations and consequently, the continuum only appears in the plot of $R_{0\rightarrow 2}$ and is absent in the plot of $R_{1\rightarrow 3}$ vs. modulation frequency, where $R_{\alpha\rightarrow \beta}$ denotes the net current from $\alpha$'th to $\beta$'th Hartree-Fock subband.

If the modulation frequency corresponds to an excitonic mode, we expect to observe transition currents between several pairs of subbands, given that the excitons are generally mixtures of p-h excitations of different subbands. For example, the red spikes in the energy absorption spectrum, which denotes an exciton, is visible in both $R_{0\rightarrow 2}$ and $R_{1\rightarrow 3}$ plots.

Finally, if an exciton lies inside a continuum, it will be damped and yield a broadened peak in the energy absorption spectrum. In this case, although the continuum is associated to a certain inter-subband transition {\em per se}, we expect to observe its trace in other inter-subband transition rate plots due to mixing with the excitonic mode. This effect is schematically shown in plots Fig.~\ref{fig:intuition}d and~\ref{fig:intuition}f. The red shaded continuum in the energy absorption spectrum denotes a $1 \rightarrow 3$ p-h continuum and the kink in the middle is indicative of an exciton lying inside it. Both the continuum and the kink appear in all of the inter-subband transition rate plots in this case.\\

In the remainder of this paper, we explore the ideas summarized above in detail. This paper is organized as follows: the microscopic model is introduced in Sec.~\ref{sec:mod} and the Hartree-Fock equation is discussed in Sec.~\ref{sec:HFintro}. The numerical results obtained by solving the Hartree-Fock equation is discussed in Sec.~\ref{sec:HFres}. Lattice modulation spectroscopy experiments are reviewed in Sec.~\ref{sec:TMS} and the calculation of energy absorption spectrum and inter-subband transition rates in the TDHF approximation is discussed in Sec.~\ref{sec:TMSmethod}. The numerical results are presented and discussed in Sec.~\ref{sec:TMSres}. The paper is concluded by a short discussion on the experimental outlook of the presented results.

\section{The Microscopic Model}\label{sec:mod}
In this section, we review the microscopic model for spinless fermions with electric dipole-dipole interactions in a one-dimensional optical trap. This model is relevant to a pancake of polar molecules prepared in a single hyperfine state, as well as to a stack of well-separated pancakes.

For concreteness, we assume that the gas is confined about the $x$-$y$ using a confining optical potential centered at $z=0$. Also, we assume that the dipoles are aligned along the $z$-axis (perpendicular to the confining plane) using a strong d.c. electric field. The Hamiltonian of the system is the sum of the optical trap potential, the kinetic energy and the electric dipole-dipole interactions. A convenient basis for the second quantized notation is one that diagonalizes the one-body part of the Hamiltonian. We choose the following basis:
\begin{equation}\label{eq:origbasis}
\langle \mathbf{r} | \alpha, \mathbf{k} \rangle = \phi_{\alpha}(z) \frac{1}{\sqrt{A}} e^{i\mathbf{k}\cdot\mathbf{x}},
\end{equation}
where $\phi_{\alpha}(z)$ is the wavefunction of the $\alpha$'th transverse mode of the trap, $A$ is the area of the trap in the $x$-$y$ plane and $\mathbf{x} = (x,y)$ is the in-plane coordinates. In this basis, the second-quantized Hamiltonian is easily found to be:
\begin{eqnarray}\label{eq:hamilt}
&&H = \sum_{\mathbf{k},\alpha} \left( \frac{\hbar^2 |\mathbf{k}|^2}{2m} + \epsilon_{\alpha}\right)c^{\dagger}_{\mathbf{k},\alpha}c^{\phantom{\dagger}}_{\mathbf{k},\alpha}\nonumber\\
&&+\frac{1}{2}\sum_{\alpha\beta;\gamma\lambda} \sum_{\kk_1\kk_2\qq} \mathcal{V}_{\alpha\beta;\gamma\lambda}(\mathbf{q})\,c^{\dagger}_{\kk_1+\qq,\alpha}c^{\dagger}_{\kk_2-\qq,\gamma}c^{\phantom{\dagger}}_{\kk_2,\lambda}c^{\phantom{\dagger}}_{\kk_1,\beta},
\end{eqnarray}
where $c_{\kk,\alpha}$ ($c^{\dagger}_{\kk,\alpha}$) annihilates (creates) a particle in the in-plane momentum state $\kk$ and $\alpha$'th subband. $\epsilon_{\alpha}$ is the zero-point energy of $\alpha$'th subband ($\alpha=0, 1,\,\ldots$). We measure the energies with respect to the zero-point energy of the lowest subband ($\alpha=0$). $\mathcal{V}_{\alpha\beta;\gamma\lambda}(\mathbf{q})$ is the Fourier transform of the effective inter-subband dipolar interaction, defined as:
\begin{eqnarray}\label{eq:vinter}
V_{\alpha\beta;\gamma\lambda}(\xx - \xx') = \int \int &&\mathrm{d}z \, \mathrm{d}z' \, \phi_{\alpha}^*(z) \, \phi_{\gamma}^*(z') \, \phi_{\lambda}(z') \, \phi_{\beta}(z) \nonumber\\&&\times\, V_{\mathrm{dip}}(z - z', \xx - \xx'),
\end{eqnarray}
where $V_{\mathrm{dip}}(z, \xx)$ is the electric dipole-dipole interaction of two particles with a center separation of $(z,\xx)$:
\begin{equation}\label{eq:vdip}
V_{\mathrm{dip}}(z, \xx) = \frac{D^2}{\left(|\xx|^2 + z^2\right)^{\frac{5}{2}}}\left(|\xx|^2 - 2 z^2\right).
\end{equation}
The intra-subband interactions are special cases of Eq.~(\ref{eq:vinter}), i.e. the intra-subband interaction in $\alpha$'th subband is given by $\mathcal{V}_{\alpha\alpha;\alpha\alpha}(\qq)$. Since the dipoles are perpendicular to the $x$-$y$ plane, the interaction matrix elements are isotropic in the in-plane momentum $\qq$.

One can always choose the transverse wavefunctions $\phi_{\alpha}(z)$ to be real and of well-defined parity for trap potentials which are symmetric about $z=0$ (harmonic trap is one example). As a result, it is easy to verify that $\mathcal{V}_{\alpha\beta;\gamma\lambda}$ is invariant under the following interchange of indices: $\alpha \leftrightarrow \beta$, $\gamma \leftrightarrow \lambda$, or $\alpha\beta \leftrightarrow \gamma\lambda$. Also, $\mathcal{V}_{\alpha\beta;\gamma\lambda}$ vanishes if $\alpha + \beta + \gamma + \lambda \equiv 1~(\mathrm{mod}~2)$. This parity-conserving behavior is due to the invariance of Eq.~(\ref{eq:vdip}) under inversion $z \rightarrow -z$, and breaks down for unaligned dipoles.\\

For simplicity, we assume that the optical trap is perfectly harmonic with a frequency $\omega_{\mathrm{trap}}$. In this case, we easily find:
\begin{eqnarray}\label{eq:ensubband}
\epsilon_{\alpha} &=& \alpha\,(\hbar\omega_{\mathrm{trap}}),\nonumber\\
\phi_{\alpha}(z) &=& \frac{1}{\sqrt{\pi^{1/2}\,\alpha!\,2^\alpha\,a_\bot}}\,H_{\alpha}(z/a_{\bot})\,e^{-z^2/2 a_{\bot}},
\end{eqnarray}
where $H_{\alpha}(z)$ is the $\alpha$'s Hermite polynomial~\cite{Arfken} and $a_\perp$ is the transverse confinement width, related to the trap frequency as $\omega_{\mathrm{trap}} = \hbar/m a_{\bot}^2$. A generating function for $\mathcal{V}_{\alpha\beta;\gamma\lambda}(\mathbf{q})$ for harmonic traps is derived in Appendix~\ref{sec:Vformula} and explicit expressions for the first few inter-subband interactions are given.

We note that in the experiments, nearly-perfect harmonic trapping can be achieved by loading the gas into a single well of a strong optical lattice. An optical lattice potential of the form $V_{\mathrm{lat.}}(z) = (\hbar^2/2 m a_{\perp}^4 k^2)\sin^2 k z$ yields a harmonic trap with frequency $\hbar/m a_{\bot}^2$ centered at $z=0$ in the limit $k \rightarrow 0$.

\section{Hartree-Fock Equations for a uniform quasi-two-dimensional Fermionic gas}\label{sec:HF}\label{sec:HFintro}
We briefly review the Hartree-Fock (HF) theory for a uniform quasi-two-dimensional fermionic gas in thermal equilibrium and apply the formalism to the system of fermionic polar molecules. The Hartree-Fock equation for quasi-two-dimensional systems is found to be substantially more difficult to solve compared to the strictly two-dimensional case due to subband hybridization.\\

We begin our treatment with the usual definition of the 1-particle thermal Green's function:
\begin{equation}
\mathcal{G}_{\mu\nu}(\kk,i\omega_n) = -\int_{0}^{\beta\hbar} \mathrm{d}\tau\,e^{i\omega_n\tau}\,\mathrm{Tr}\big[\hat{\rho}_\mathrm{G}\,c^{\phantom{\dagger}}_{\kk,\mu}(\tau)c^{\dagger}_{\kk,\nu}(0)\big],
\end{equation}
where $\beta=1/k_B T$, $T$ is the temperature, $i\omega_n = (2n+1)\pi/\beta$ is the fermionic Matsubara frequency, $\hat{\rho}_\mathrm{G} = e^{-\beta H}/\mathrm{Tr}[e^{-\beta H}]$ is the grand-canonical statistical weighting operator and $c^{(\dagger)}_{\kk,\mu}(\tau) = e^{\tau H/\hbar}\,c^{(\dagger)}_{\kk,\mu}\,e^{-\tau H/\hbar}$ is the imaginary-time Heisenberg fermion annihilation (creation) operator.  In the presence of interactions, the non-interacting subband indices no longer remain good quantum numbers due to hybridization and consequently, $\mathcal{G}_{\mu\nu}(\kk,i\omega_n)$ is expected to have non-zero off-diagonal elements.

The Hartree-Fock approximation for the Green's function is given by the following diagrammatic Dyson's equation~\cite{FetterWalecka}:
\begin{eqnarray}\label{diag:HF}
&&\parbox{30pt}{
  \begin{fmfgraph*}(30,70)
  \fmfbottom{i}
  \fmftop{o}	
  \fmf{fermion,width=thick,label=${}_{q}$,l.s=right}{i,o}
  \fmffreeze
  \fmfiv{l=${}_{\beta}$,l.a=0,l.d=6}{vloc(__i)}
  \fmfiv{l=${}_{\alpha}$,l.a=0,l.d=6}{vloc(__o)}
  \end{fmfgraph*}}
=
\parbox{40pt}{
 \begin{fmfgraph*}(30,70)
  \fmfbottom{i}
  \fmftop{o}	
  \fmf{fermion,width=thin,label=${}_{q}$,l.s=right}{i,o}
  \fmffreeze
  \fmfiv{l=${}_{\beta}$,l.a=0,l.d=6}{vloc(__i)}
  \fmfiv{l=${}_{\alpha}$,l.a=0,l.d=6}{vloc(__o)}
  \end{fmfgraph*}}
+
\parbox{60pt}{
  \begin{fmfgraph*}(60,70)
  \fmfforce{(0.3w,0.0h)}{i}
  \fmfforce{(0.3w,1.0h)}{o}  	
  \fmfforce{(0.9w,0.5h)}{v2}
  \fmfforce{(0.65w,0.5h)}{v3}
  \fmfdot{v1}
  \fmfdot{v3}
  \fmf{fermion,width=thick,label=${}_{q}$,l.s=left}{i,v1}
  \fmf{fermion,width=thin,label=${}_{q}$,l.s=left}{v1,o}
  \fmf{zigzag,width=thin,tension=0}{v1,v3}
  \fmf{fermion,width=thick,label=${}_{k'}$,label.side=right,right}{v2,v3}
  \fmf{plain,width=thick,right}{v3,v2}
  \fmffreeze
  \fmfiv{label=${}_{\beta}$,l.a=180,l.d=4}{vloc(__i)}
  \fmfiv{label=${}_{\nu}$,l.a=-60,l.d=8}{vloc(__v1)}
  \fmfiv{label=${}_{\mu}$,l.a=60,l.d=8}{vloc(__v1)}
  \fmfiv{label=${}_{\alpha}$,l.a=180,l.d=4}{vloc(__o)}
  \fmfiv{label=${}_{\lambda}$,l.a=100,l.d=8}{vloc(__v3)}
  \fmfiv{label=${}_{\gamma}$,l.a=-100,l.d=8}{vloc(__v3)}
\end{fmfgraph*}}
+
\parbox{60pt}{
  \begin{fmfgraph*}(60,70)
  \fmfforce{(0.3w,0.0h)}{i}
  \fmfforce{(0.3w,1.0h)}{o}
  \fmf{fermion,width=thick,label=${}_{q}$,l.s=right}{i,v1}
  \fmf{fermion,width=thick,label=${}_{k'}$,l.s=left,tension=0.7}{v1,v2}
  \fmf{fermion,width=thin,label=${}_{q}$,l.s=right}{v2,o}
  \fmf{zigzag,right=0.75,width=thin,label=${}_{q-k'}$,tension=0}{v1,v2}
  \fmfdot{v1}
  \fmfdot{v2}
  \fmffreeze
  \fmfiv{label=${}_{\beta}$,l.a=180,l.d=4}{vloc(__i)}
  \fmfiv{label=${}_{\alpha}$,l.a=180,l.d=4}{vloc(__o)}
  \fmfiv{label=${}_{\gamma}$,l.a=150,l.d=4}{vloc(__v1)}
  \fmfiv{label=${}_{\nu}$,l.a=-150,l.d=4}{vloc(__v1)}
  \fmfiv{label=${}_{\mu}$,l.a=150,l.d=4}{vloc(__v2)}
  \fmfiv{label=${}_{\lambda}$,l.a=-150,l.d=4}{vloc(__v2)}
  \end{fmfgraph*}}
\end{eqnarray}
where the thin and thick fermion lines denote non-interacting and interacting Green's functions respectively. The diagram yields the following equation:
\begin{eqnarray}\label{eq:mbHF}
\GG_{\alpha\beta}(\kk,i\omega_n) & = & \GG^0_{\alpha\beta}(\kk,i\omega_n) + \GG^0_{\alpha\mu}(\kk,i\omega_n)\Sigma^{\star}_{\mu\nu}(\kk)\nonumber\\&&\times\,\GG_{\nu\beta}(\kk,i\omega_n),
\end{eqnarray}
where the proper self-energy $\Sigma^{\star}_{\mu\nu}(\kk)$ is the sum of the direct and exchange diagrams:
\begin{eqnarray}\label{eq:mbsen}
\Sigma^{\star}_{\mu\nu}(\kk) &=& \frac{1}{\beta}\sum_{i\omega'_n}\int \frac{\mathrm{d}^2\kk'}{(2\pi)^2}\left[\mathcal{V}_{\mu\nu;\gamma\lambda}(0)-\mathcal{V}_{\mu\lambda;\gamma\nu}(\kk-\kk')\right]\nonumber\\&&\times\,\GG_{\lambda\gamma}(\kk',i\omega'_n)\,e^{i\omega'_n 0^+}.
\end{eqnarray}
Summation over repeated indices is assumed throughout this paper. The non-interacting thermal Green's function is given by:
\begin{equation}
\GG^0_{\alpha\beta}(\kk,i\omega_n) = \frac{\delta_{\alpha\beta}}{i\omega_n - \xi_{\kk,\alpha}}, \qquad \xi_{\kk,\alpha}=\frac{|\kk|^2}{2m}+\epsilon_{\alpha}-\mu.
\end{equation}
In the absence of spontaneous symmetry breaking of the inversion symmetry (which may happen in the strongly interacting regime and is beyond the scope of this paper), the interactions only mix subbands of the same parity due to the symmetries of the inter-subband interaction matrix elements mentioned in the previous section. Investigating Eqs.~(\ref{eq:mbHF}) and~(\ref{eq:mbsen}) shows that $\GG_{\alpha\beta} = 0$ if $\alpha + \beta \equiv 1~(\mathrm{mod}~2)$ and therefore, $\GG_{\alpha\beta}$ will have the following matrix structure in the non-interacting subband indices:
\begin{equation}\label{eq:matrixstruct}
\GG = \left(\begin{array}{cccccc}
\GG_{00} & 0 & \GG_{02} & 0 & \GG_{04} & \ldots\\
0 & \GG_{11} & 0 &  \GG_{13} & 0 & \ldots\\
\GG_{02} & 0 & \GG_{22} & 0 &  \GG_{24} & \ldots\\
0 & \GG_{13} & 0 & \GG_{33} & 0 & \ldots\\
\GG_{04} & 0 & \GG_{24} & 0 &  \GG_{44} & \ldots\\
\vdots & \vdots & \vdots & \vdots & \vdots & \ddots \\
\end{array}\right).
\end{equation}
In order to solve the Hartree-Fock equation at finite temperatures, one must carry out the Matsubara summations appearing in Eq.~(\ref{eq:mbsen}). The summation can be easily done once the dependence of the Green's function on the Matsubara frequency is explicitly known. In fact, expressing the interacting Green's function its diagonal basis reveals its explicit dependence on the Matsubara frequency. Treated as a matrix equation, Eq.~(\ref{eq:mbHF}) formally yields:
\begin{eqnarray}\label{eq:matrix}
\GG &=& \left[\left(\GG^0\right)^{-1} - \Sigma^{\star}\right]^{-1}\nonumber\\
&=& \left[i\omega_n\mathbf{I} - \overline{\Sigma}^{\star}\right]^{-1},
\end{eqnarray}
where $\overline{\Sigma}^{\star}_{\alpha\beta}(\kk) = \Sigma^{\star}_{\alpha\beta}(\kk) + \xi^0_{\kk,\alpha}\delta_{\alpha\beta}$, i.e. the proper self-energy including the kinetic energy contribution. It is easy to see that $\overline{\Sigma}^{\star}(\kk)$ is a symmetric matrix and therefore, there exists a real orthonormal basis in which it is diagonal and has real eigenvalues. Let $U(\kk)$ be the unitary transformation that diagonalizes $\overline{\Sigma}^{\star}(\kk)$:
\begin{eqnarray}\label{eq:diagonal}
\overline{\Sigma}^{\star}(\kk) &=& U(\kk)\,\Xi(\kk)\,U^{T}(\kk),\nonumber\\
\Xi(\kk) &=& \mathrm{diag}\{\tilde{\xi}_1(\kk),~\tilde{\xi}_2(\kk),~\ldots\},
\end{eqnarray}
where $\{\tilde{\xi}_\alpha(\kk)\}$ are the eigenvalues.
The same transformation clearly diagonalizes the interacting Green's function:
\begin{eqnarray}\label{eq:mbnewG}
\tilde{\GG}_{\alpha\beta}(\kk,i\omega_n) &\stackrel{\text{def}}{=}& \left[U(\kk)^{T}\GG U(\kk)\right]_{\alpha\beta} = \left[i\omega_n\mathbf{I} - \Xi(\kk)\right]^{-1}_{\alpha\beta} \nonumber\\&=& \frac{\delta_{\alpha\beta}}{i\omega_n - \tilde{\xi}_{\alpha}(\kk)}.
\end{eqnarray}
As promised, the Matsubara frequency summation appearing in Eq.~(\ref{eq:mbsen}) can be evaluated with ease in the new basis:
\begin{eqnarray}
&&\hspace{-20pt}\frac{1}{\beta}\sum_{i\omega'_n}\GG_{\lambda\gamma}(\kk',i\omega'_n)\,e^{i\omega'_n 0^+}\nonumber\\
&&=\frac{1}{\beta}\sum_{i\omega'_n} \sum_\rho U_{\lambda\rho}(\kk')U_{\gamma\rho}(\kk')\,\frac{e^{i\omega'_n 0^+}}{i\omega'_n-\tilde{\xi}_{\rho}(\kk')}\nonumber\\
&&=\sum_\rho U_{\lambda\rho}(\kk')\,n^F[\tilde{\xi}_{\rho}(\kk')]\,U_{\gamma\rho}(\kk').
\end{eqnarray}
where $n^F(x) = \left(e^{\beta x}+1\right)^{-1}$ is the Fermi occupation function, resulting from the summation over fermionic Matsubara frequencies~\cite{FetterWalecka}. Plugging this result into Eq.~(\ref{eq:mbHF}), we get an explicit self-consistent equation for the proper self-energy:
\begin{eqnarray}\label{eq:finalhf}
\Sigma^{\star}_{\mu\nu}(\kk) &=& \int \frac{\mathrm{d}^2\kk'}{(2\pi)^2}\left[\mathcal{V}_{\mu\nu;\gamma\lambda}(0)-\mathcal{V}_{\mu\lambda;\gamma\nu}(\kk-\kk')\right]U_{\lambda\rho}(\kk')\nonumber\\&&\times\, U_{\gamma\rho}(\kk')n^F\left[\xi_{\rho}(\kk')\right].
\end{eqnarray}
It is understood that $U(\kk)$ and $\xi_\rho(\kk)$ are implicit functions of $\Sigma^{\star}(\kk)$, defined in Eq.~(\ref{eq:diagonal}).\\

\begin{figure*}[t!]
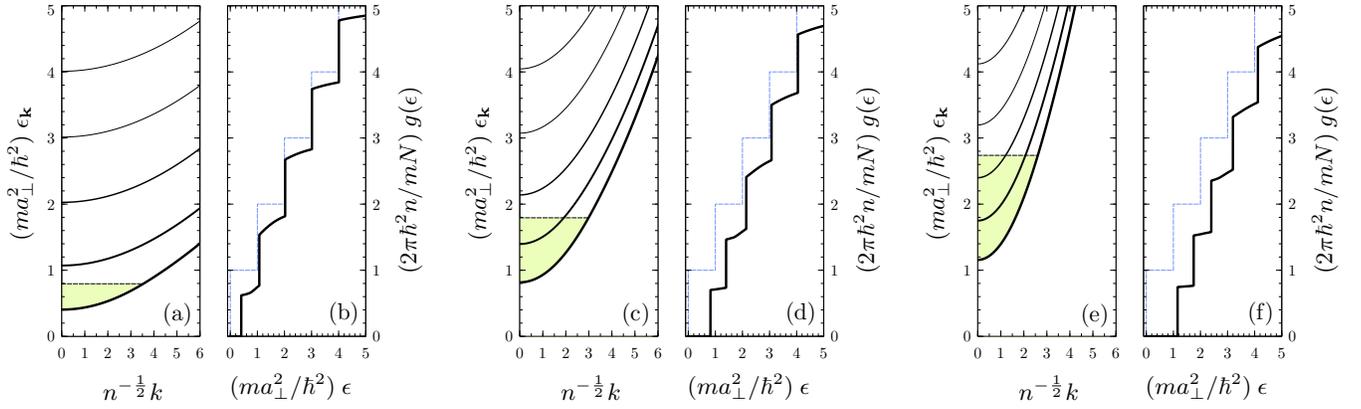

\subfigure{\begin{lpic}{figs/band_struct1(5.5cm,5.5cm)}
\lbl[bl]{10,63,90;$(m a_{\perp}^2/\hbar^2)~\epsilon_{\kk}$}
\lbl[bl]{151,51,90;$(2\pi \hbar^2 n / m N)~g(\epsilon)$}
\lbl[bl]{35,5;$n^{-\frac{1}{2}}k$}
\lbl[bl]{81,5;$(m a_{\perp}^2/\hbar^2)~\epsilon$}
\end{lpic}}\hspace{10pt}
\subfigure{\begin{lpic}{figs/band_struct2(5.5cm,5.5cm)}
\lbl[bl]{10,63,90;$(m a_{\perp}^2/\hbar^2)~\epsilon_{\kk}$}
\lbl[bl]{151,51,90;$(2\pi \hbar^2 n / m N)~g(\epsilon)$}
\lbl[bl]{35,5;$n^{-\frac{1}{2}}k$}
\lbl[bl]{81,5;$(m a_{\perp}^2/\hbar^2)~\epsilon$}
\end{lpic}}\hspace{10pt}
\subfigure{\begin{lpic}{figs/band_struct3(5.5cm,5.5cm)}
\lbl[bl]{10,63,90;$(m a_{\perp}^2/\hbar^2)~\epsilon_{\kk}$}
\lbl[bl]{151,51,90;$(2\pi \hbar^2 n / m N)~g(\epsilon)$}
\lbl[bl]{35,5;$n^{-\frac{1}{2}}k$}
\lbl[bl]{81,5;$(m a_{\perp}^2/\hbar^2)~\epsilon$}
\end{lpic}}
\caption{The energy dispersion of the first five Hartree-Fock subbands (panels a, c, and e) and the density of states (DOS) $g(\epsilon)$ (panels b, d and f) for $r_d = 1.0$, $T/T^* = 0.1$. (a) and (b): $\sqrt{n}a_{\perp} = 0.2$ (c) and (d): $\sqrt{n}a_{\perp} = 0.4$ (e) and (f): $\sqrt{n}a_{\perp} = 0.6$. The DOS in the absence of interactions is also shown for comparison (blue dashed lines).}
\label{fig:band}
\end{figure*}

It is clear from the preceding discussion that the orthogonal transformation $U(\kk)$ also defines the mean-field single-particle states in the presence of interactions. We identify $\tilde{\xi}_{\alpha}(\kk)$ as the energy dispersion of $\alpha$'th Hartree-Fock subband and define the Hartree-Fock fermion annihilation (creation) operators as:
\begin{equation}\label{eq:HFop}
\tilde{c}^{(\dagger)}_{\kk,\alpha}(\tau) = \sum_{\mu} U_{\mu\alpha}(\kk) \, c^{(\dagger)}_{\kk,\mu}(\tau).
\end{equation}
It is straight-forward to show in light of Eq.~(\ref{eq:mbnewG}) that $\tilde{\GG}_{\mu\nu}(\kk,i\omega_n)$ can be identically defined using the Hartree-Fock fermion operators:
\begin{equation}
\tilde{\GG}_{\mu\nu}(\kk,i\omega_n) = -\int_{0}^{\beta\hbar} \mathrm{d}\tau\,e^{i\omega_n\tau}\,\mathrm{Tr}\big[\hat{\rho}_{\mathrm{G}}^{\mathrm{HF}}\tilde{c}^{\phantom{\dagger}}_{\kk,\mu}(\tau)\tilde{c}^{\dagger}_{\kk,\nu}(0)\big],
\end{equation}
where $\hat{\rho}_{\mathrm{G}}^{\mathrm{HF}}$ is the grand-canonical operator defined in terms of the Hartree-Fock decoupled Hamiltonian.\\

At this point, it is also useful to define the effective interaction between Hartree-Fock quasiparticles. The evaluation of response functions, which is our goal in the next section, is more natural in this basis. Expressing the interaction part of the Hamiltonian in terms of Hartree-Fock fermion operators, one can easily read off the renormalized interaction between incoming particles in Hartree-Fock subbands $\beta$ and $\lambda$ with momenta $\kk_1$ and $\kk_2$, scattering to subbands $\alpha$ and $\gamma$ with momenta $\kk_1+\qq$ and $\kk_2-\qq$ respectively:
\begin{eqnarray}\label{eq:HFintervertex}
&&\tilde{\mathcal{V}}_{\alpha\beta;\gamma\lambda}(\kk_1,\kk_2,\qq) = \sum_{\alpha'\beta'\gamma'\lambda'}\mathcal{V}_{\alpha'\beta';\gamma'\lambda'}(\qq)~U_{\alpha'\alpha}(\kk_1+\qq) \nonumber\\&&\quad \times\,U_{\gamma'\gamma}(\kk_2-\qq) \, U_{\lambda'\lambda}(\kk_2) \, U_{\beta'\beta}(\kk_1).
\end{eqnarray}
Note that the renormalized interaction is no longer just a function of the momentum transfer, but also depends on the individual momenta of the scattering quasiparticles.

\subsection{Results: {\em renormalized bands}}\label{sec:HFres}

\begin{figure*}
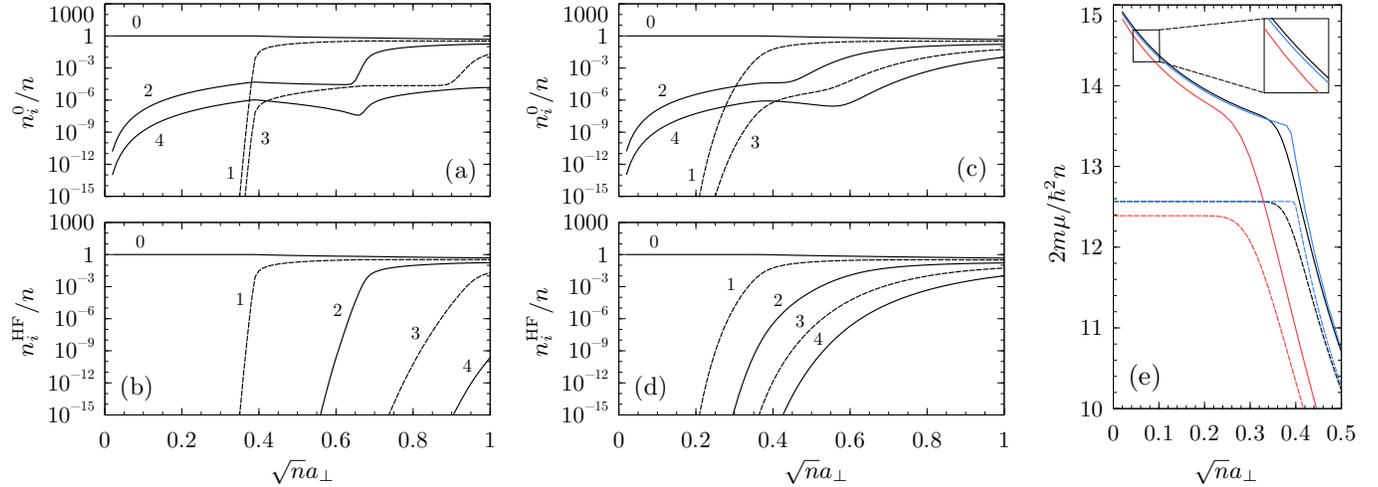

\hspace{-15pt}\subfigure{\begin{lpic}{figs/new_sb_dens1(6.5cm,)}
\lbl[bl]{10,111,90;$n^0_i/n$}
\lbl[bl]{10,41,90;$n^{\mathrm{HF}}_i/n$}
\lbl[bl]{81,2;$\sqrt{n}a_{\perp}$}
\end{lpic}}\hspace{5pt}
\subfigure{\begin{lpic}{figs/new_sb_dens2(6.5cm,)}
\lbl[bl]{10,111,90;$n^0_i/n$}
\lbl[bl]{10,41,90;$n^{\mathrm{HF}}_i/n$}
\lbl[bl]{81,2;$\sqrt{n}a_{\perp}$}
\end{lpic}}\hspace{10pt}
\subfigure{\begin{lpic}[l(0mm),r(-30mm),t(0mm),b(0mm)]{figs/mu1(,6.5cm)}
\lbl[bl]{7,70,90;$2m\mu/\hbar^2 n$}
\lbl[bl]{45,2;$\sqrt{n}a_{\perp}$}
\end{lpic}}
\caption{Subband mixing and renormalization of the chemical potential. Panels (a) and (b) show the fractional population of non-interacting subbands and Hartree-Fock subbands respectively, as a function of transverse confinement width $a_{\perp}$ at constant dipolar interaction strength $r_d = 0.05$ and temperature $T=0.1\,T^*\simeq 0.008\,T_F^{(0)}$. Panels (c) and (d) are the same quantities at a higher temperature $T=T^*\simeq 0.08\,T_F^{(0)}$. Panel (e) shows the chemical potential as a function of transverse confinement width $a_{\perp}$ at $T=0.1\,T^*\simeq 0.008\,T_F^{(0)}$ (solid blue), $T=1.0\,T^*\simeq 0.08\,T_F^{(0)}$ (solid black) and $T=2.0\,T^*\simeq 0.16\,T_F^{(0)}$ (solid red). The dashed lines denote the chemical potential in the absence of interactions for comparison. Notice the non-monotonic ordering of the chemical potential vs. temperature in the inset (refer to the text for discussions).}
\label{fig:dens}
\end{figure*}

\begin{figure}
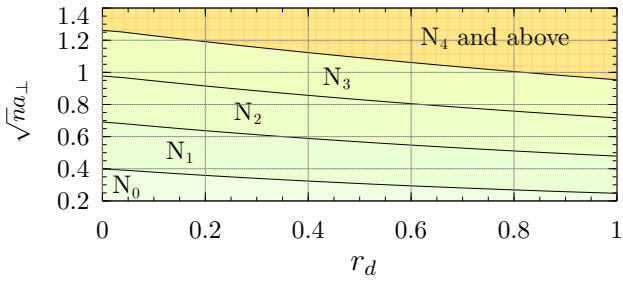

\begin{lpic}[l(-2mm),r(0mm),t(0mm),b(-42mm)]{figs/band_transition_lines(8cm,)}
\lbl[bl]{2,115,90;$\sqrt{n}a_{\perp}$}
\lbl[bl]{82,82,0;\large $r_d$}
\end{lpic}
\caption{The quasi-two-dimensional liquid phases of fermionic polar molecules at $T=0$. The abbreviations $\mathrm{N}_0$, $\mathrm{N}_1$, etc stand for normal liquid phases populating up to the zeroth, first, ... Hartree-Fock subbands respectively.}
\label{fig:phasediag}
\end{figure}

In this section, we present the numerical results obtained by solving the Hartree-Fock equation. The numerical method is described in Appendix~\ref{sec:HFnum} in detail. The temperatures are reported in the units of $T^*$, which is defined as:
\begin{eqnarray}
T^* &=& \hbar^2 n/2 m k_B.
\end{eqnarray}
Note that $T^*$ ua related to $T_F^{(0)}$, the Fermi temperature of a two-dimensional non-interacting Fermi gas, as:
\begin{equation}
T^* = T_F^{(0)}/4\pi \simeq 0.08\,T_F^{(0)}.
\end{equation}

Fig.~\ref{fig:band} shows the energies of the first five Hartree-Fock subbands (panels a, c and e) and their corresponding density of states (DOS) (panels b, d and f) for $r_d = 1.0$, $T/T^*~=~0.1$ and for three different transverse confinement widths $\sqrt{n}a_{\perp} = 0.2$, $0.4$ and $0.6$. Deviations from the non-interacting quadratic energy dispersions can also be observed in the DOS plots: the DOS of a quasi-two-dimensional non-interacting Fermi gas has a uniform staircase structure (shown in Fig.~\ref{fig:band}b,~\ref{fig:band}d and~\ref{fig:band}f as blue dashed lines for reference). In the presence of interactions, we find that (1) the DOS plot starts at a finite energy, meaning that the zero-point energy of the lowest subband is lifted, (2) the DOS plot no longer has flat regions (due to deviations from quadratic), (3) the energy spacing between the jumps, which correspond to the spacing between the zero-point energies of the subbands, become non-uniform, and (4) the DOS of a non-interacting gas is always larger than the DOS of the interacting gas, which is associated to the long-range repulsive nature of dipolar interactions in the studied confined geometry.

At zero temperature, the occupation of higher subbands is only due to Pauli exclusion and the Fermi occupation function is sharp. Therefore, at any given density, trap strength and interaction strength, only a finite number of Hartree-Fock subbands are fully or partially occupied. There, a phase diagram can be obtained for the system at $T=0$ as a function of trap and dipolar interaction strengths (Fig.~\ref{fig:phasediag}). It is noticed that at fixed transverse confinement width $a_{\perp}$ and density $n$, stronger interactions result in occupation of higher subbands. This behavior can be understood in light of the reduction of DOS due to interaction. The energy and fractional occupation of the subbands were found to be continuous across the phase boundaries and therefore, the transitions are continuous. The same transitions has been reported to be first-order for quasi-two-dimensional electron gas~\cite{Goni2002}.

In order to see the hybridization of non-interacting subbands, we have plotted the fractional density of non-interacting and Hartree-Fock subbands in Fig.~\ref{fig:dens}a and~\ref{fig:dens}b respectively, as a function of transverse confinement width and at fixed $r_d=0.05$ and temperature $T/T^* = 0.1$.  Fig.~\ref{fig:dens}c and ~\ref{fig:dens}c show the same quantities at a higher temperature $T/T^* = 1.0$. It is observed that larger transverse confinement widths (i.e. weaker traps) naturally results in occupation of higher subbands. Hybridization is clearly noticeable by comparing Fig.~\ref{fig:dens}a and~\ref{fig:dens}b: occupation of the lowest Hartree-Fock subband amounts to occupation of several non-interacting subbands of even parity, $0, 2, 4, ~\ldots\,$, with decreasing weights. A consequence of subband mixing is anomalous population inversion, i.e. the 2nd non-interacting subband is populated before the 1st non-interacting subband, due to hybridization with 0th, 4th, \ldots~subbands.

Finally, Fig.~\ref{fig:dens}e shows the renormalized chemical potential as a function of transverse confinement width at three different temperatures (solid lines). We have also plotted the chemical potential in the absence of interactions for reference (dashed lines). It is noticed that the renormalized chemical potential is always greater than its non-interacting value, which is again due to reduction of the DOS in the presence of interactions. An interesting observation is the non-monotonic behavior of the chemical potential as a function of temperature, which is clearly noticeable in the inset plot of Fig.~\ref{fig:dens}e. As a consequence, the isothermal compressibility of the interacting gas, $\kappa_T = n^{-2} (\partial n / \partial \mu)_T$ also turns out to have a non-monotonic behavior as a function of temperature. This behavior has been reported earlier~\cite{Kestner2010} and is confirmed by our calculations. Intuitively, a small rise in temperature will result in thermal excitation of states just below the Fermi level. Having larger momenta, the thermally excited quasiparticles experience stronger interactions and decrease DOS at the Fermi level. Consequently, the chemical potential has to be increased in order to compensate for the reduced DOS. At higher temperatures, the effect is suppressed as the states above the Fermi level are significantly populated and a smaller chemical potential is required to keep the number of particles constant.

\section{Probing the inter-subband excitations in lattice modulation spectroscopy experiments}\label{sec:TMS}
The single particle and collective excitations of a strictly two-dimensional gas of polar fermionic molecules has been recently studied by several authors~\cite{Bruun2008,Ronen2010,Yamaguchi2010}. The two-dimensional gas corresponds to the single-subband limit of a quasi-two-dimensional gas and can be achieved by increasing the trap frequency. At zero tilt angle, i.e. when the dipoles are aligned perpendicular to the two-dimensional plane of the trap, the dipole-dipole interactions are effectively repulsive. The dipolar interactions are also short-range and regular in the long wavelength limit. Therefore, such systems are expected have the same qualitative properties as He-3, which is described well by Fermi liquid theory of neutral systems~\cite{NozieresPines}. The elementary excitations of such systems consists of single particle excitations and the zero sound.

As the higher subbands are populated, we encounter a new class of elementary excitations. Analogous to the strictly two-dimensional case, the inter-subband excitations also come in two flavors: inter-subband quasiparticle-like excitations and p-h bound states (excitons).

The inter-subband excitations can be experimentally detected using the technique of lattice modulation spectroscopy~\cite{Stoeferle2004,Iucci2006,Kollath2006,Sensarma2009,Tokuno2011} which was originally introduced and realized in cold atomic systems in order to study the Mott insulator and superfluid phases of the simulated Hubbard model. This method relies on the high controllability of the optical lattices which allows rapid modulation of the optical potential. Introducing a weak oscillatory modulation to the amplitude of the optical potential, one stimulates the transition of particles from lower subbands to higher subbands. The energy of the inter-subband excitations must match the modulation frequency for the transition to occur. Therefore, one directly measures the excitation energy in such experiments and no tedious calibration of parameters is needed.

If one is interested to measure the energy absorbed in the process of lattice modulation, one allows the system to re-thermalize after the lattice modulation pulse. All optical potentials (including the polarizing d.c. electric field in the case of polar molecules) are suddenly switched off then and the density profile of the gas is measured after an interval of ballistic expansion. The change in the width of the central peak in the momentum profile can be taken as a measure of the absorbed energy. By carrying out this procedure for a range of frequencies, one obtains the inter-subband excitation spectrum of the system. We refer to this procedure as {\em energy-resolved measurement} for brevity.

Another quantity which is often measured in lattice modulation spectroscopy experiments is the change in the number of particles in each subband. We refer to such experiments as {\em band-resolved measurements} for brevity. In such experiments, one avoids the re-thermalization of the gas after the modulation pulse. Instead, the gas is allowed to expand ballistically quickly after the modulation pulse and the transverse momentum profile of the particles is measured. The particles are resolved into subbands by fitting the measured density profile to a weighted sum of the density profiles calculated from the transverse wavefunctions. Band-resolved measurements can be used to study the mechanism of energy absorption.

\subsection{The Perturbative Formulation of Lattice Modulation Spectroscopy Experiments}\label{sec:TMSmethod}
In this section, we describe a theoretical framework for predicting the results of modulation spectroscopy experiments of polar molecules. As mentioned in Sec.~\ref{sec:mod}, we are interested in the case of well separated layers, where one can neglect the inter-layer couplings and just focus on transitions within a single layer. For concreteness, we focus on the pancake at $z=0$. Upon introducing the amplitude modulation, the optical potential takes the following time-dependent form:
\begin{eqnarray}\label{eq:mod}
V_{\mathrm{lat.}}(z,t) & = & \left[V_0 + V_1 f(t)\right]\,\sin^2(kz)\nonumber\\
& \simeq & \frac{1}{2}\,m\,\omega_{\mathrm{trap}}^2\,z^2 + \xi\,f(t)\,z^2 / a_{\perp}^2,
\end{eqnarray}
where $\omega_{\mathrm{trap}} = k \sqrt{2V_0/m}$, $a_{\perp} = \sqrt{\hbar/(m\omega_{\mathrm{trap}})}$, $\xi = (\hbar\,\omega_{\mathrm{trap}}/2V_0)V_1$ and $f(t)$ is the shape of the amplitude modulation pulse. We have expanded the optical potential to quadratic order about its minimum at $z=0$. The second quantized form of the lattice modulation potential, i.e. the second term in Eq.~(\ref{eq:mod}), can be written as:
\begin{equation}\label{eq:Hpert}
\hat{V}_{\mathrm{mod}}(t) = \xi\,f(t)\,\sum_{\kk,\alpha\beta} \mathbf{T}_{\alpha\beta}\, c^{\dagger}_{\kk\alpha}c^{\phantom{\dagger}}_{\kk\beta},
\end{equation}
where $\mathbf{T}_{\alpha\beta} = a_{\perp}^{-2}\int \mathrm{d}z\,\phi^*_{\alpha}(z) \, z^2 \, \phi_{\beta}(z)$. For a harmonic trap, $\mathbf{T}_{\alpha\beta}$ is given by:
\begin{eqnarray}\label{eq:t}
\mathbf{T}_{\alpha\beta} &=& \frac{1}{2}\Big( \sqrt{(\alpha+1)(\alpha+2)}\,\delta_{\alpha,\beta-2}\nonumber\\
&& + \sqrt{\alpha(\alpha-1)}\,\delta_{\alpha,\beta+2} + (2\alpha+1)\,\delta_{\alpha\beta}\Big).
\end{eqnarray}
It is easier for subsequent derivations to express the perturbation Hamiltonian in terms of Hartree-Fock fermion operators:
\begin{equation}\label{eq:Hpert2}
\hat{V}_{\mathrm{mod}}(t) = \xi\,f(t)\,\sum_{\kk,\alpha\beta} \tilde{\mathbf{T}}_{\alpha\beta}(\kk) \, \tilde{n}_{\alpha\beta}(\kk),
\end{equation}
where:
\begin{eqnarray}
\tilde{n}_{\alpha\beta}(\kk) &=& \tilde{c}^{\dagger}_{\kk\alpha} \tilde{c}^{\phantom{\dagger}}_{\kk\beta},\nonumber\\
\tilde{\mathbf{T}}_{\alpha\beta}(\kk) &=& \sum_{\alpha'\beta'} U_{\alpha'\alpha}(\kk)U_{\beta'\beta}(\kk)t_{\alpha'\beta'}.
\end{eqnarray}
We note that both $\mathbf{T}_{\alpha\beta}$ and $\tilde{\mathbf{T}}_{\alpha\beta}$ are symmetric is the subband indices. It is clear from Eqs.~(\ref{eq:Hpert}) and~(\ref{eq:t}) that this perturbation stimulates an inter-subband p-h excitation between subbands whose index differ by $\pm 2$. Note that we have tacitly assumed that the amplitude of the modulation is low (i.e. $\xi \ll \hbar \omega_{\mathrm{trap}}$ in Eq.~\ref{eq:mod}) and  neglected the small anharmonic terms such as $\sim \cos(\omega t)\,z^4$, which stimulate transitions between subbands whose index differ by $\pm 4$ as well.\\

For low-amplitude modulations and short modulation pulse durations, the disturbance in the equilibrium state of the system is expected to be negligible and therefore, we can treat the modulation potential in perturbation theory. In the first order approximation, the response of any dynamical variable of system to an oscillatory external field is also purely oscillatory. Therefore, the linear response vanishes on average and one needs to consider processes which are at least second order in the external fields in order to describe transport phenomena such as energy absorption and inter-band transitions.\\

We focus on energy-resolved measurements first. A straightforward calculation yields the following expression for $\dot{E}(t)\equiv\partial_t \mathrm{Tr}[\hat{\rho}(t)\hat{H}] $ in the second-order perturbation theory:
\begin{align}\label{eq:Edot0}
\dot{E}(t) &= 2\,\mathfrak{Re}\int_{-\infty t}\mathrm{d}t'\,\langle \hat{V}_{\mathrm{mod}}(t)\,\hat{H}\,\hat{V}_{\mathrm{mod}}(t')\nonumber\\
&\hspace{120pt}-\hat{H}\,\hat{V}_{\mathrm{mod}}(t)\,\hat{V}_{\mathrm{mod}}(t')\rangle\nonumber\\
&=i\xi^2\,f(t) \int_{-\infty}^{\infty}\mathrm{d}t'\,f(t')\,\partial_t\,\mathcal{T}(t-t'),
\end{align}
where:
\begin{equation}\label{eq:defT}
\mathcal{T}(t)=\sum_{\kk_1,\alpha\beta}\sum_{\kk_2,\gamma\lambda}\mathcal{T}_{\alpha\beta;\gamma\lambda}(\kk_1,\kk_2,t),
\end{equation}
and:
\begin{equation}\label{eq:defT1}
\mathcal{T}_{\alpha\beta;\gamma\lambda}(\kk_1,\kk_2,t)=\tilde{\mathbf{T}}_{\alpha\beta}(\kk_1)\,\tilde{\mathbf{T}}_{\gamma\lambda}(\kk_2)
\,\tilde{\Pi}^{R}_{\alpha\beta;\gamma\lambda}(\kk_1,\kk_2;t),
\end{equation}
in which $\tilde{\Pi}^{R}_{\alpha\beta;\gamma\lambda}(\kk_1,\kk_2;t)$ is the retarded inter-subband polarization insertion:
\begin{equation}
\tilde{\Pi}^{R}_{\alpha\beta;\gamma\lambda}(\kk_1,\kk_2;t) =-i\theta(t)\, \mathrm{Tr}\left\{\hat{\rho}_0\left[\tilde{n}_{\beta\alpha}(\kk_1,t),\tilde{n}_{\gamma\lambda}(\kk_2,0)\right]\right\}.
\end{equation}

\begin{figure}[t]
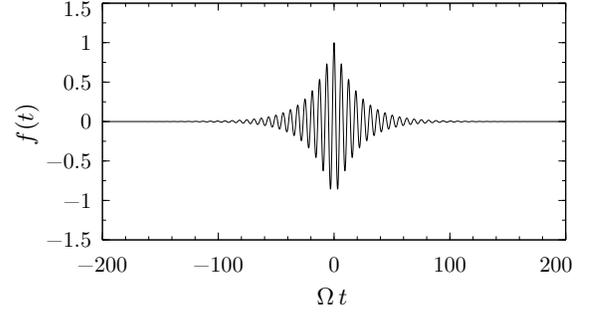

\begin{lpic}[l(0mm),r(0mm),t(0mm),b(2mm)]{figs/pulse(7cm,)}
\lbl[bl]{-2,39,90;$f(t)$}
\lbl[bl]{78,-7,0;$\Omega\,t$}
\end{lpic}
\caption{Plot of the a.c. pulse for $\eta/\Omega = 0.05$. The band-resolved measurement is done at $t=0$.}
\label{fig:pulse}
\end{figure}

In the experiments, the lattice modulation pulse is switched on and off smoothly. For concreteness, we assume an exponentially switched-on pulse:
\begin{equation}\label{eq:pulse}
f(t)=e^{-\eta|t|}\,\cos(\Omega\,t),
\end{equation}
where $\eta$ is the switching rate and $\Omega$ is the frequency of the a.c. modulation. We chose the exponentially switched-on pulse (shown in Fig.~\ref{fig:pulse}) since it yields simple and transparent analytical expressions. It is shown in Appendix~\ref{app:Edot} that the energy absorption rate for such a pulse is given by:
\begin{equation}\label{eq:Edot}
\dot{E}\big|_{\Omega,\eta} = -\frac{1}{4}\,\Omega\,\mathfrak{Im}\left[\mathcal{T}(\Omega + i\eta)\right] + \mathcal{O}(\hbar\,\xi^2\, \eta\,\omega^{-1}_{\mathrm{trap}}).
\end{equation}
Note that in light of Eqs.~(\ref{eq:defT}) and~(\ref{eq:defT1}), $\mathcal{T}(\omega)$, the Fourier transform of $\mathcal{T}(t)$, can be identified as the self-energy correction of the optical lattice photons with energy $\hbar\omega$ coupled to the fermionic molecules. Thus, Eq.~(\ref{eq:Edot}) can be simply interpreted as the energy of the absorbed photons multiplied by their decay rate, i.e. the imaginary part of their self-energy.\\
It is evident from Eq.~(\ref{eq:defT}) that the fundamental quantity to be evaluated is $\tilde{\Pi}^{R}_{\alpha\beta;\gamma\lambda}(\kk_1,\kk_2;t)$, i.e. the dynamical inter-subband polarization insertion with zero net momentum transfer. We evaluate this quantity in the conserving time-dependent Hartree-Fock (TDHF) approximation~\cite{BaymKadanoff1961}, also known as Generalized Random Phase Approximation (GRPA)~\cite{NozieresPines} or Random Phase Approximation with Exchange (RPAE)~\cite{Lipparini1995}. Once the inter-subband polarizations are found, one can easily evaluate the energy absorption rate using Eqs.~(\ref{eq:defT}),~(\ref{eq:defT1}) and~(\ref{eq:Edot}). We carry out the diagrammatic calculations in the imaginary-time formalism. The retarded polarization appearing in Eq.~(\ref{eq:defT1}) can be found by the standard procedure of analytical continuation of the Matsubara frequency to the upper complex frequency half-plane: 
\begin{eqnarray}
\tilde{\Pi}_{\alpha\beta;\gamma\lambda}(\kk_1,\kk_2;\Omega) = \tilde{\Pi}_{\alpha\beta;\gamma\lambda}(\kk_1,\kk_2;i\nu_n\rightarrow\,\Omega).\nonumber\\
\end{eqnarray}

Evaluating polarization diagrams in the TDHF approximation amounts to summing the direct and exchange scatterings between the p-h pairs to all orders in each p-h loop. In other words, the full polarization diagram is the sum of all {\em ring} diagrams with {\em ladder}-like vertex corrections~\cite{BaymKadanoff1961}. The successive iterations of following Bethe-Salpeter equation generates all such contributions:
\begin{eqnarray}\label{eq:TDHF}
&&\hspace{-40pt}\parbox{70pt}{
  \begin{fmfgraph*}(70,60)
  \fmfleft{i1}
  \fmfleft{i2}
  \fmfright{o1}
  \fmfright{o2}
  \fmfforce{(3.5,45)}{i1}
  \fmfforce{(3.5,15)}{i2}
  \fmfforce{(66.5,45)}{o1}
  \fmfforce{(66.5,15)}{o2}
  \fmfforce{(20,45)}{v1}
  \fmfforce{(20,15)}{v2}
  \fmfpolyn{filled=10,label=${}_{{\Pi}_{\alpha\beta;\gamma\lambda}}$,tension=1}{v}{4}
  \fmf{dashes_arrow,label=${}_{k_1,,\beta}$,label.side=left}{i1,v1}
  \fmf{dashes_arrow,label=${}_{k_1,,\alpha}$,label.side=left}{v2,i2}
  \fmf{dashes_arrow,label=${}_{k_2,,\lambda}$,label.side=left}{v4,o1}
  \fmf{dashes_arrow,label=${}_{k_2,,\gamma}$,label.side=left}{o2,v3}
  \end{fmfgraph*}}=
  \begin{tabular}{c}
  	${\scriptstyle\delta_{\alpha\gamma}}$\\
 	${\scriptstyle\delta_{\beta\lambda}}$\\
 	${\scriptstyle\delta_{k_1k_2}}$
  \end{tabular}
\parbox{20pt}{
  \begin{fmfgraph*}(20,60)
  \fmfforce{(0.05w,0.75h)}{i1}
  \fmfforce{(0.05w,0.25h)}{i2}
  \fmfforce{(0.95w,0.75h)}{o1}
  \fmfforce{(0.95w,0.25h)}{o2}
  \fmf{fermion,width=thick,label=${}_{k_1,,\beta}$,label.side=left}{i1,o1}
  \fmf{fermion,width=thick,label=${}_{k_1,,\alpha}$,label.side=left}{o2,i2}
  \end{fmfgraph*}} +
\parbox{70pt}{
  \begin{fmfgraph*}(90.5,60)
  \fmfleft{i1}
  \fmfleft{i2}
  \fmfleft{vv1}
  \fmfleft{vv2}
  \fmfright{o1}
  \fmfright{o2}
  \fmfforce{(3.5,45)}{i1}
  \fmfforce{(3.5,15)}{i2}
  \fmfforce{(20,45)}{vv1}
  \fmfforce{(20,15)}{vv2}
  \fmfforce{(87,45)}{o1}
  \fmfforce{(87,15)}{o2}
  \fmfforce{(40.5,45)}{v1}
  \fmfforce{(40.5,15)}{v2}
  \fmfpolyn{filled=10,label=${}_{{\Pi}_{\alpha\beta;\gamma\lambda}}$,tension=1}{v}{4}
  \fmf{dashes_arrow,label=${}_{k_1,,\beta}$,label.side=left}{i1,vv1}
  \fmf{dashes_arrow,label=${}_{k_1,,\alpha}$,label.side=left}{vv2,i2}
  \fmf{fermion,label=${}_{k',,\nu}$,label.side=left,width=thick}{vv1,v1}
  \fmf{fermion,label=${}_{k',,\mu}$,label.side=left,width=thick}{v2,vv2}
  \fmf{dashes_arrow,label=${}_{k_2,,\lambda}$,label.side=left}{v4,o1}
  \fmf{dashes_arrow,label=${}_{k_2,,\gamma}$,label.side=left}{o2,v3}
  \fmf{dbl_zigzag}{vv1,vv2}
  \end{fmfgraph*}}
\end{eqnarray}
where the double zigzag line represents the combined direct and exchange interaction between Hartree-Fock quasiparticles:
\begin{eqnarray}
\tilde{\mathcal{U}}_{\alpha\beta;\mu\nu}^{(\mathrm{d+e})}(\kk,\kk') &=& \parbox{40pt}{
  \begin{fmfgraph*}(40,60)
  \fmfleft{ii1}
  \fmfleft{ii2}
  \fmfright{o1}
  \fmfright{o2}
  \fmfforce{(0.05w,0.75h)}{ii1}
  \fmfforce{(0.05w,0.25h)}{ii2}
  \fmfforce{(0.50w,0.75h)}{i1}
  \fmfforce{(0.50w,0.25h)}{i2}
  \fmfforce{(0.95w,0.75h)}{o1}
  \fmfforce{(0.95w,0.25h)}{o2}
  \fmf{dbl_zigzag}{i1,i2}
  \fmf{dashes_arrow}{ii1,i1}
  \fmf{dashes_arrow}{i1,o1}
  \fmf{dashes_arrow}{o2,i2}
  \fmf{dashes_arrow}{i2,ii2}
  \fmffreeze
  \fmfiv{label=${}_{k,,\beta}$,l.a=90}{vloc(__ii1)}
  \fmfiv{label=${}_{k,,\alpha}$,l.a=-90}{vloc(__ii2)}
  \fmfiv{label=${}_{k',,\nu}$,l.a=90}{vloc(__o1)}
  \fmfiv{label=${}_{k',,\mu}$,l.a=-90}{vloc(__o2)}
  \end{fmfgraph*}} = \parbox{40pt}{
  \begin{fmfgraph*}(40,60)
  \fmfleft{ii1}
  \fmfleft{ii2}
  \fmfright{o1}
  \fmfright{o2}
  \fmfforce{(0.05w,0.75h)}{ii1}
  \fmfforce{(0.05w,0.25h)}{ii2}
  \fmfforce{(0.50w,0.75h)}{i1}
  \fmfforce{(0.50w,0.25h)}{i2}
  \fmfforce{(0.95w,0.75h)}{o1}
  \fmfforce{(0.95w,0.25h)}{o2}
  \fmf{zigzag}{i1,i2}
  \fmf{dashes_arrow}{ii1,i1}
  \fmf{dashes_arrow}{i1,o1}
  \fmf{dashes_arrow}{o2,i2}
  \fmf{dashes_arrow}{i2,ii2}
  \fmffreeze
  \fmfiv{label=${}_{k,,\beta}$,l.a=90}{vloc(__ii1)}
  \fmfiv{label=${}_{k,,\alpha}$,l.a=-90}{vloc(__ii2)}
  \fmfiv{label=${}_{k',,\nu}$,l.a=90}{vloc(__o1)}
  \fmfiv{label=${}_{k',,\mu}$,l.a=-90}{vloc(__o2)}
  \end{fmfgraph*}} +
\parbox{70pt}{
  \begin{fmfgraph*}(70,60)
  \fmfleft{ii1}
  \fmfleft{ii2}
  \fmfright{o1}
  \fmfright{o2}�
  \fmfforce{(0.05w,0.75h)}{ii1}
  \fmfforce{(0.05w,0.25h)}{ii2}
  \fmfforce{(0.35w,0.50h)}{i1}
  \fmfforce{(0.65w,0.50h)}{i2}
  \fmfforce{(0.95w,0.75h)}{o1}
  \fmfforce{(0.95w,0.25h)}{o2}
  \fmf{dashes_arrow,left=0.2}{ii1,i1}
  \fmf{dashes_arrow,left=0.2}{i2,o1}
  \fmf{dashes_arrow,left=0.2}{o2,i2}
  \fmf{dashes_arrow,left=0.2}{i1,ii2}
  \fmf{zigzag}{i1,i2}
  \fmffreeze
  \fmfiv{label=${}_{k,,\beta}$,l.a=30}{vloc(__ii1)}
  \fmfiv{label=${}_{k,,\alpha}$,l.a=-30}{vloc(__ii2)}
  \fmfiv{label=${}_{k',,\mu}$,l.a=-130}{vloc(__o2)}
  \fmfiv{label=${}_{k',,\nu}$,l.a=155}{vloc(__o1)}
  \end{fmfgraph*}}\nonumber\\
&=&\tilde{\mathcal{V}}_{\nu\beta;\alpha\mu}(\kk,\kk',\kk'-\kk) - \tilde{\mathcal{V}}_{\alpha\beta;\nu\mu}(\kk,\kk',0).\nonumber\\
\end{eqnarray}
The negative sign of the direct interaction is due to the extra Fermion loop it introduces. The diagrammatic Eq.~(\ref{eq:TDHF}) yields:
\begin{multline}\label{eq:TDHF1}
\tilde{\Pi}_{\alpha\beta;\gamma\lambda}(\bar{k}_1,\bar{k}_2;i\nu_n) = \tilde{\mathcal{G}}_{\alpha}(\kk_1,i\omega_{n_1}+i\nu_n)\tilde{\mathcal{G}}_{\beta}(\kk_1,i\omega_{n_2}) \\
\Bigg[\delta_{\alpha\gamma}\delta_{\beta\lambda}\delta_{\kk_1 \kk_2}
-\frac{1}{\beta}\sum_{i\omega'_n}\int\frac{\mathrm{d}^2\kk'}{(2\pi)^2}\,\tilde{\mathcal{U}}_{\alpha\beta;\mu\nu}^{(\mathrm{d}+\mathrm{e})}(\kk_1,\kk')\\
\times\,\tilde{\Pi}_{\mu\nu;\gamma\lambda}(\bar{k}',\bar{k}_2;i\nu_n)\Bigg],
\end{multline}
where $\bar{k}_1 = (\kk_1,i\omega_{n_1})$, $\bar{k}_2 = (\kk_2,i\omega_{n_2})$ and $\bar{k}' = (\kk',i\omega'_{n})$. We note that if we had formulated the problem in the non-interacting basis, the Green's functions would have been non-diagonal in the subband indices and the resulting integral equation would have had additional intermediate subband index summations.

Summing both sides of Eq.~(\ref{eq:TDHF1}) over the Matsubara frequencies $i\omega_{n_1}$ and $i\omega_{n_2}$ and analytically continuing $i\nu_n \rightarrow \Omega$, we get:
\begin{eqnarray}\label{eq:integPI}
&&\tilde{\Pi}_{\alpha\beta;\gamma\lambda}(\kk_1,\kk_2;\Omega) = \tilde{\Pi}^{(0)}_{\alpha\beta}(\kk_1;\Omega)\,\Bigg[\delta_{\alpha\gamma}\delta_{\beta\lambda}\delta_{\kk_1\kk_2}\nonumber\\
&&\quad-\int\frac{\mathrm{d}^2\kk'}{(2\pi)^2} \, \tilde{\mathcal{U}}_{\alpha\beta;\mu\nu}^{(\mathrm{d}+\mathrm{e})}(\kk_1,\kk') \, \tilde{\Pi}_{\mu\nu;\gamma\lambda}(\kk',\kk_2;\Omega)\Bigg],
\end{eqnarray}
where, $\tilde{\Pi}^{(0)}_{\alpha\beta}(\kk;\Omega)$, the bare inter-subband polarization with zero net momentum transfer, is defined as:
\begin{eqnarray}\label{eq:barepol}
\tilde{\Pi}^{(0)}_{\alpha\beta}(\kk;\Omega) &=& \frac{1}{\beta}\sum_{i\omega_n}\tilde{\mathcal{G}}_{\alpha}(\kk;i\omega_n+i\nu_n)\tilde{\mathcal{G}}_{\beta}(\kk;i\omega_n)\bigg|_{i\nu_n\rightarrow\Omega}\nonumber\\
&=&\frac{n_{\beta}(\kk)-n_{\alpha}(\kk)}{\Omega + \tilde{\xi}_{\beta}(\kk) - \tilde{\xi}_{\alpha}(\kk)}.
\end{eqnarray}
The poles of $\tilde{\Pi}^{(0)}_{\alpha\beta}(\kk;\Omega)$ correspond to $\alpha \rightarrow \beta$ p-h excitation energies.

Eq.~(\ref{eq:integPI}) is an integral equation for the inter-subband polarization. However, it is more suitable for our purpose to formulate an integral equation for the following auxiliary function, in which some of the summations appearing in Eq.~(\ref{eq:defT}) is already carried out:
\begin{equation}\label{eq:defh}
\tilde{h}_{\alpha\beta}(\kk_1;\Omega) = \frac{\sum_{\kk_2,\gamma\lambda}\tilde{\Pi}_{\alpha\beta;\gamma\lambda}(\kk_1,\kk_2;\Omega)\,\tilde{\mathbf{T}}_{\gamma\lambda}(\kk_2)}{\tilde{\Pi}^{(0)}_{\alpha\beta}(\kk_1;\Omega)}.
\end{equation}
Using Eq.~(\ref{eq:integPI}), one easily finds:
\begin{eqnarray}\label{eq:integh}
&&\tilde{h}_{\alpha\beta}(\kk_1;\Omega) = \tilde{\mathbf{T}}_{\alpha\beta}(\kk_1)\nonumber\\
&&\quad-\int\frac{\mathrm{d}^2\kk'}{(2\pi)^2}\,\tilde{\mathcal{U}}_{\alpha\beta;\mu\nu}^{(\mathrm{d}+\mathrm{e})}(\kk_1,\kk')\,\tilde{\Pi}^{(0)}_{\mu\nu}(\kk';\Omega)\,\tilde{h}_{\mu\nu}(\kk';\Omega).\nonumber\\
\end{eqnarray}
Iterating the above integral equation,
\begin{center}
\includegraphics[scale=0.75]{figs/feyn-h}
\end{center}
one identifies $\tilde{h}_{\alpha\beta}$ as the effective inter-subband transition matrix. The presence of p-h scatterings to all orders suggests that $\tilde{h}_{\alpha\beta}(\kk;\Omega)$ will have poles at excitonic energies. Note that in contrast to Eq.~(\ref{eq:integPI}), the bare polarizations (which have integrable singularities) appear under integrals in the above equation. Therefore, provided that its Fredholm determinant is non-vanishing, Eq.~(\ref{eq:integh}) yields smooth solutions. However, the Fredholm determinant may vanish for an isolated set of frequencies, corresponding to the excitons. We use this criteria for locating the excitons.\\

Once $h_{\alpha\beta}(\kk;\Omega)$ is found, $\mathcal{T}(\Omega)$ can be readily evaluated using Eq.~(\ref{eq:defh}):
\begin{eqnarray}\label{eq:Tph}
\mathcal{T}(\Omega) &=&\sum_{\alpha\beta}\sum_{\kk_1}\tilde{\mathbf{T}}_{\alpha\beta}(\kk_1)\,\tilde{\Pi}^{(0)}_{\alpha\beta}(\kk_1;\Omega)\,\tilde{h}_{\alpha\beta}(\kk_1;\Omega).
\end{eqnarray}
The numerical procedure used to solve the integral equation for $\tilde{h}_{\alpha\beta}$ is described in detail in Appendix~\ref{sec:TMSnum}.\\

In the TDHF approximation, the excitons that lie outside of p-h continuums are undamped and have infinitely long lifetimes. Therefore, the excitonic poles appear infinitesimally below the real frequency axis. Thus, according to Eq.~(\ref{eq:Edot}), such undamped excitons yield sharp Dirac delta-like peaks in the energy absorption spectrum in the adiabatic pulse switching limit ($\eta\rightarrow 0$). The spectral weight of an exciton can be found using the analyticity of $\mathcal{T}(\omega)$ in the upper complex frequency half-plane. Assuming that that $\mathcal{T}(\omega)$ is singular at $\omega = \omega_0 - i0^+$, one can find the spectral weight associated to it using the Kramers-Kronig relations:
\begin{equation}\label{eq:colwght}
W\big|_{\omega=\omega_0} = \frac{\pi \xi^2}{4}\,\omega_0\lim_{\omega\rightarrow\omega_0}\Big\{(\omega-\omega_0)\,\mathfrak{Re}\big[\mathcal{T}(\omega)\big]\Big\}.
\end{equation}\\

Finally, we address the subtler problem of evaluating the inter-subband transition rates. The subtlety originates from the fact that the number of particles in each subbands is not conserved separately by the microscopic Hamiltonian. The presence of inter-subband interaction matrix elements such as $\mathcal{V}_{\gamma\alpha;\gamma\beta}$ implies that $[\hat{N}_{\gamma},\hat{H}] \neq 0$, where $\hat{N}_{\gamma}$ is the total number operator of particles in $\gamma$'th subband. We note that this result holds for Hartree-Fock number operators as well. As a consequence, the inter-subband excitations generated by the lattice modulation term are subject to non-equilibrium dynamics as the system evolves. Ultimately, one needs to use the Schwinger-Keldysh formalism and non-equilibrium Green's functions in order to calculate the transition rates in such situations. We note that this additional difficulty is not present in theories in which the number of each species is conserved separately. In those cases, the transition rates can be evaluated using the equilibrium formalism in the second-order perturbation theory~\cite{Mahan}.

We quote the final result of the non-equilibrium analysis and leave the technical details for future works. In brief, our strategy is to solve the non-equilibrium Dyson's equation up to second order in the external field. The inter-subband transition rates can be extracted from the second-order corrections to the non-equilibrium Green's functions. For the exponentially switched-on pulse, we find that the total number of particles transferred from the Hartree-Fock subband $\alpha$ to $\beta$ for times $t\leq 0$ can be written as:
\begin{equation}\label{eq:Iab}
\Delta \tilde{N}_{\alpha \rightarrow \beta}(t)\Big|_{t \leq 0} = \frac{e^{-2\eta |t|}}{4\eta}
\big|\tilde{h}_{\beta\alpha}(\Omega+i\eta)\big|^2
\mathfrak{Im}\left[\tilde{\Pi}^{(0)}_{\alpha\beta}(\Omega+i\eta)\right].
\end{equation}
We have neglected the oscillatory terms in the above equation. Such contributions are the second harmonics of the external a.c. field which naturally arise in the second-order perturbation theory. In the limit $\eta/\Omega \ll 1$, such terms are much smaller than the retained term. Moreover, they yield no steady transition current on average due to their oscillatory nature.

The effective duration of the exponentially switched-on pulse from $t=-\infty$ to $t=0$, when the measurement takes place, is $\Delta t \sim \eta^{-1}$. We define the effective inter-subband transition rate as:  
\begin{equation}\label{eq:Rab}
\tilde{\mathcal{R}}_{\alpha \rightarrow \beta} \equiv \eta\,\Delta \tilde{N}_{\alpha \rightarrow \beta}(0).
\end{equation}
For weakly interacting systems, subband hybridization is a negligible effect (see Fig.~\ref{fig:dens}; the fractional hybridization is $\sim 10^{-6}$ for $r_d = 0.05$) and the transition rates between Hartree-Fock subbands are virtually the same as the transition rates between the non-interacting subbands, i.e $\Delta \tilde{N}_{\alpha \rightarrow \beta}(t) \approx \Delta N_{\alpha \rightarrow \beta}(t)$ and $\tilde{\mathcal{R}}_{\alpha \rightarrow \beta} \approx \mathcal{R}_{\alpha \rightarrow \beta}$.\\

We conclude this section by a short discussion on the conditions for validity of the time-dependent Hartree-Fock theory. We begin with the subtler case of evaluating the rate of change of non-conserved quantities. Note that conserved and non-conserved quantities are defined with respect to the Hamiltonian in the absence of external fields. Generally, we expect the prediction of the Hartree-Fock theory to become less reliable for longer pulse durations. In particular, the study of the adiabatic limit must be avoided in this approximation. The reason is that the collision integrals are neglected in the Hartree-Fock approximation and consequently, the resulting picture lacks relaxation and rethermalization mechanisms. Hence, the theory yields unphysically long-lived excitations (with the only exception of bound states lying inside p-h continuums) and leads to erroneous predictions for transition rates. Therefore, one must make sure that the lifetime of excitations are larger than the pulse duration. For the exponentially switched-on pulse of Eq.~(\ref{eq:pulse}), this condition leads to the following criteria: 
%
\begin{equation}\label{eq:relaxcrit}
\langle\Gamma^{\mathrm{rel.}}_{\alpha}\rangle < 2\eta \quad (\mathrm{for~all~}\alpha),
\end{equation}
where $\langle\Gamma^{\mathrm{rel.}}_{\alpha}\rangle$ is the average relaxation rate of an excitation in $\alpha$'th subband. An estimate of $\langle\Gamma^{\mathrm{rel.}}_{2}\rangle$ for quasiparticle-like excitations is provided in Appendix~\ref{app:relax}. We will use Eq.~(\ref{eq:relaxcrit}) and the results of Appendix~\ref{app:relax} in order to choose an appropriate value for $\eta$ in the next section. We note that in experiments with reactive polar molecules, another constraint on the pulse duration results from the reduced lifetime of molecules in the higher subbands. In practice, $\eta$ must be chosen to be larger than both the relaxation rate and the molecule loss rate.

Finally, we remark that the reliability the Hartree-Fock predictions for the evolution of conserved quantities, such as the expectation value of energy, is expected to be unaffected by long pulse durations. The reason is that the rate of energy absorption processes, i.e. creation of excitations, is essentially determined by the equilibrium states of the system and their occupation, which is assumed to be only slightly disturbed by the external perturbation during the experiment. In other words, relaxation processes only affect the evolution of non-conserved quantities.

\subsection{Results}\label{sec:TMSres}
In this section, we present the results obtained by evaluating Eq.~(\ref{eq:Edot}) and~(\ref{eq:Iab}) numerically. We keep the first five subbands in the numerical calculations. The results are given in terms of the following dimensionless and intensive quantities:
\begin{align}
\dot{\varepsilon}& \equiv (2\hbar N^{-1} \xi^{-2})\,\dot{E},\nonumber\\
w & \equiv (2\hbar N^{-1} \xi^{-2} \omega^{-1}_{\mathrm{trap}})\,W,\nonumber\\
r_{\alpha\rightarrow\beta} & \equiv (2\hbar^2 N^{-1} \xi^{-2})\,\Omega\, \mathcal{R}_{\alpha\rightarrow\beta},\nonumber\\
\bar{\Omega} &\equiv \Omega/\omega_{\mathrm{trap}}.
\end{align}
Based on the remarks mentioned at the end of the previous section, one expects the energy absorption spectrum obtained in experiments done at a finite switching rate to be essentially a broadened version of those obtained in the adiabatic limit. In fact, the appearance of $\mathcal{T}(\Omega+i\eta)$ in Eq.~(\ref{eq:Edot}) and the analyticity of $\mathcal{T}(\omega)$ in the upper complex frequency half-plane is a rigorous justification of this claim. In light of this observation, we have evaluated the energy absorption rates in the adiabatic limit for the clarity of presentation. In this limit, the excitonic peaks will be Dirac delta-like and one can easily differentiate between the excitons and the p-h continuums.

\subsubsection{Energy Absorption Rates}
\begin{figure}[h]
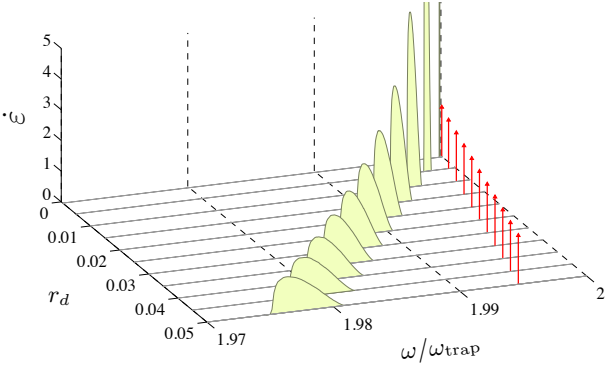

\subfigure{\begin{lpic}[l(-5mm),r(0mm),t(0mm),b(-5mm)]{figs/EAR_dip2(9.0cm,)}
\lbl[bl]{17,90,90;\large $\dot{\varepsilon}$}
\lbl[bl]{25,30,0;$r_d$}
\lbl[bl]{143,10,7;$\omega/\omega_{\mathrm{trap}}$}
\end{lpic}}\hspace{20pt}
\caption{Energy absorption rate plots for various dipolar interaction strengths at fixed $\sqrt{n}a_{\perp} = 0.2$ and $T = 0.1\,T^* \simeq 0.008 \,T_F^{(0)}$. In all of the shown cases, the spectrum consists of a single p-h excitation continuum (yellow shaded continuum) and a single anti-bound exciton (red spikes). The exciton captures $99.74\%$ of the spectral weight for the range of $r_d$ shown in the figure.}
\label{fig:EARdip2}
\end{figure}

\begin{figure}[t]
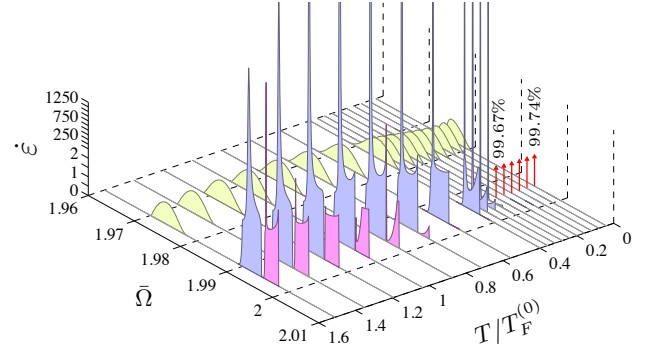

\subfigure{\begin{lpic}[l(0mm),r(0mm),t(0mm),b(-5mm)]{figs/EAR_temp(9.0cm,)}
\lbl[bl]{12,80,90;\large $\dot{\varepsilon}$}
\lbl[bl]{45,30,0;$\bar{\Omega}$}
\lbl[bl]{160,15,20;$T/T_F^{(0)}$}
\lbl[bl]{180,83,90;$\scriptstyle 99.74\%$}
\lbl[bl]{168,79,90;$\scriptstyle 99.67\%$}
\end{lpic}}
\caption{Energy absorption rate plots for various temperatures at fixed $\sqrt{n}a_{\perp} = 0.2$ and $r_d = 0.05$ (achievable in the current experiments of the group at JILA). The excitons appear as red spikes and their fractional spectral weights are shown above them. At low temperatures, the spectrum consists of a single p-h continuum corresponding to $0 \rightarrow 2$ transitions (yellow shaded continuums) and a single exciton (red spikes). At higher temperatures, more p-h excitation continuums appear continuously due to thermal occupation of higher subbands. The blue and violet shaded continuums correspond to $1 \rightarrow 3$ and $2 \rightarrow 4$ transitions respectively. The excitonic modes capture more than $99\%$ of the spectral weight, whether they are outside or inside p-h continuums.}
\label{fig:EARtemp}
\end{figure}

\begin{figure*}[t]
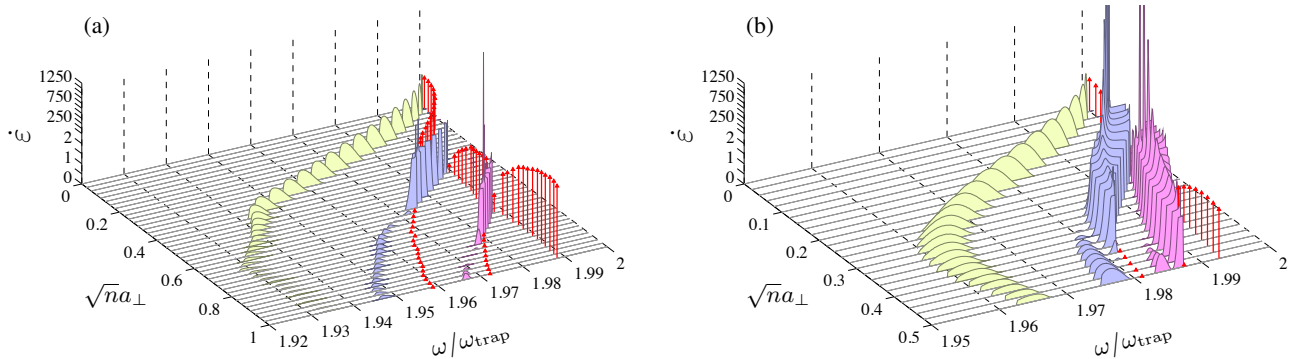

\subfigure{\begin{lpic}[l(0mm),r(0mm),t(0mm),b(-5mm)]{figs/EAR_abot(9.0cm,)}
\lbl[bl]{12,85,90;\large $\dot{\varepsilon}$}
\lbl[bl]{30,30,0;$\sqrt{n}a_{\perp}$}
\lbl[bl]{147,12,15;$\omega/\omega_{\mathrm{trap}}$}
\lbl[bl]{30,120,0;(a)}
\end{lpic}}\hspace{-10pt}
\subfigure{\begin{lpic}[l(0mm),r(0mm),t(0mm),b(-5mm)]{figs/EAR_abot_high(9.0cm,)}
\lbl[bl]{12,85,90;\large $\dot{\varepsilon}$}
\lbl[bl]{30,30,0;$\sqrt{n}a_{\perp}$}
\lbl[bl]{147,12,15;$\omega/\omega_{\mathrm{trap}}$}
\lbl[bl]{30,120,0;(b)}
\end{lpic}}
\caption{Energy absorption rate plots for various transverse confinement widths at fixed $r_d = 0.05$. (a) $T = 0.1\,T^* \simeq 0.008 \,T_F^{(0)}$, (b) $T = 10\,T^* \simeq 0.8 \,T_F^{(0)}$. The excitons appear as red spikes. For strong transverse confinements (small $a_{\perp}$), the spectrum consists of a single p-h continuum corresponding to $0 \rightarrow 2$ transitions (yellow shaded continuums) and a single exciton (red spikes). For weaker transverse confinements (larger $a_{\perp}$), more p-h excitation continuums appear continuously due to the reduced energy gap between the subbands, corresponding to transitions $1 \rightarrow 3$ (blue), $2 \rightarrow 4$ (violet), etc. The length of the spikes indicate their dimensionless spectral weight, $w$. The excitonic modes capture more than $99\%$ of the spectral weight, whether they are outside or inside p-h continuums.}
\label{fig:EARabot}
\end{figure*}


We start with the simpler case of a strong trap at low temperatures, where only the zeroth subband is populated. Fig.~\ref{fig:EARdip2} shows the energy absorption spectrum for a range of weak dipolar interactions $r_d = 0.001,~0.005,~0.010,~\ldots,~0.050$ at a constant transverse confinement width $\sqrt{n}a_{\perp} = 0.2$ and temperature $T = 0.1\,T^* \simeq 0.008 \,T_F^{(0)}$. It is noticed that the spectrum consists of a single p-h continuum, corresponding to $0 \rightarrow 2$ transitions, and a single exciton. The energy of the exciton lies above the p-h continuum and therefore, it is an anti-bound p-h pair. The anti-binding nature of the excitonic mode is a consequence of the anisotropic dipole-dipole interactions. The exciton captures $99.74\% \pm 0.01\%$ of the spectral weight, leaving only $0.26\% \pm 0.01\%$ for the continuum. These ratios were found to be independent of $r_d$ in the studied range, though, may depend on $a_{\perp}$ and $T$.\\

Fig.~\ref{fig:EARtemp} shows the energy absorption spectrum for various temperatures in the range $0.04 < T/T_F^{(0)} < 1.60$ at a constant transverse confinement width $\sqrt{n}a_{\perp} = 0.2$ and dipolar interaction strength $r_d = 0.05$. It is found that at low temperatures, the spectrum consists of a single exciton and a single p-h continuum corresponding to $0 \rightarrow 2$ transitions (yellow regions). As the temperature is increased, the spectral weight of the excitonic mode slightly drops from $99.74\% \pm 0.01\%$ to $99.67\% \pm 0.01\%$. At this point, the p-h continuum associated to $1 \rightarrow 3$ transitions (shown as blue) becomes visible due to thermal population of the first excited subband. The exciton also merges into the $1 \rightarrow 3$ continuum. At higher temperatures, more p-h continuums appear continuously due to thermal population of higher subbands. It is also noticed that the $1 \rightarrow 3$ exciton, which appears initially on the edge of $1 \rightarrow 3$ continuum (the peak at the end of $1 \rightarrow 3$ continuum, visible for $T/T_F^{(0)} \simeq 0.4$ to $\sim 1.0$), moves out of continuum and merges into $2 \rightarrow 4$ continuum (visible for $T/T_F^{(0)} \gtrsim 1.0$). We expect that all of the excitons and continuums converge to a single peak at $\omega = 2\omega_{\mathrm{trap}}$ at temperatures well beyond quantum degeneracy, where all interaction effects are masked due to thermal fluctuations. The shift of $0 \rightarrow 2$ continuum towards $\omega = 2\omega_{\mathrm{trap}}$ and its shrinking for $T>T_F^{(0)}$ agrees with this speculation.\\

Fig.~\ref{fig:EARabot} shows the energy absorption spectrum for various transverse confinement widths at a fixed dipolar interaction strength $r_d = 0.05$ for two different temperatures $T = 0.1\,T^* \simeq 0.008 \,T_F^{(0)}$ (a) and $T = 10\,T^* \simeq 0.8 \,T_F^{(0)}$ (b). The low temperature and high temperature results have strikingly similar features. At strong confinements (small $a_{\perp}$), the spectrum consists of a single exciton and a single p-h continuum. Like before, the exciton captures more than $99\%$ of the spectral weight. Upon relaxing the trap (increasing $a_{\perp}$), the energy gap between the subbands is reduced and the higher subbands will be populated, resulting in appearance of more p-h continuums and excitonic modes associated to each continuum. In the low temperature case (Fig.~\ref{fig:EARabot}a), the excitons lie outside of continuums in most of the plots while at higher temperatures, the continuums are broadened and the excitons lie inside the continuums in most cases. Had we included higher subbands in the calculations, the bare exciton appearing in Fig.~\ref{fig:EARabot}b for $\sqrt{n}a_\perp>0.3$ would have appeared inside the $3\rightarrow 5$ p-h continuum.

We find that as soon as a new exciton appears (upon increasing $a_\perp$), it continuously captures the spectral weight of the exciton below it as the trap is relaxed further. Thus, one may question the nature of the excitons, i.e. whether they are associated to certain inter-subband transitions or have a mixed nature and inherit the properties of the previous excitons which were depleted of spectral weight. We address this question in the next section where we discuss inter-subband transition rates.

Another interesting finding is the appearance of Fano line-shape~\cite{Fano1961} in parts of the spectrum. This phenomenon is most easily noticeable in the $1 \rightarrow 3$ continuums for $\sqrt{n}a_\perp > 0.34$ in Fig.~\ref{fig:EARabot}b. The {\em missing} spectral weight inside the continuum is associated to the interference of the p-h excitations and the excitonic mode just above it. 

It is worthy of mention that despite the fact that all of the results presented so far belong to the weakly interacting regime ($r_d \leq 0.05$), one finds that the presence of interactions dramatically modifies the inter-subband excitation spectrum. In the absence of interactions, all of the structures shown in Figs.~\ref{fig:EARdip2},~\ref{fig:EARtemp} and~\ref{fig:EARabot} disappear, leaving a single peak at $\omega = 2\omega_{\mathrm{trap}}$ (see Fig.~\ref{fig:intuition}).

\subsubsection{Inter-subband Transition Rates}
\begin{figure*}[t]
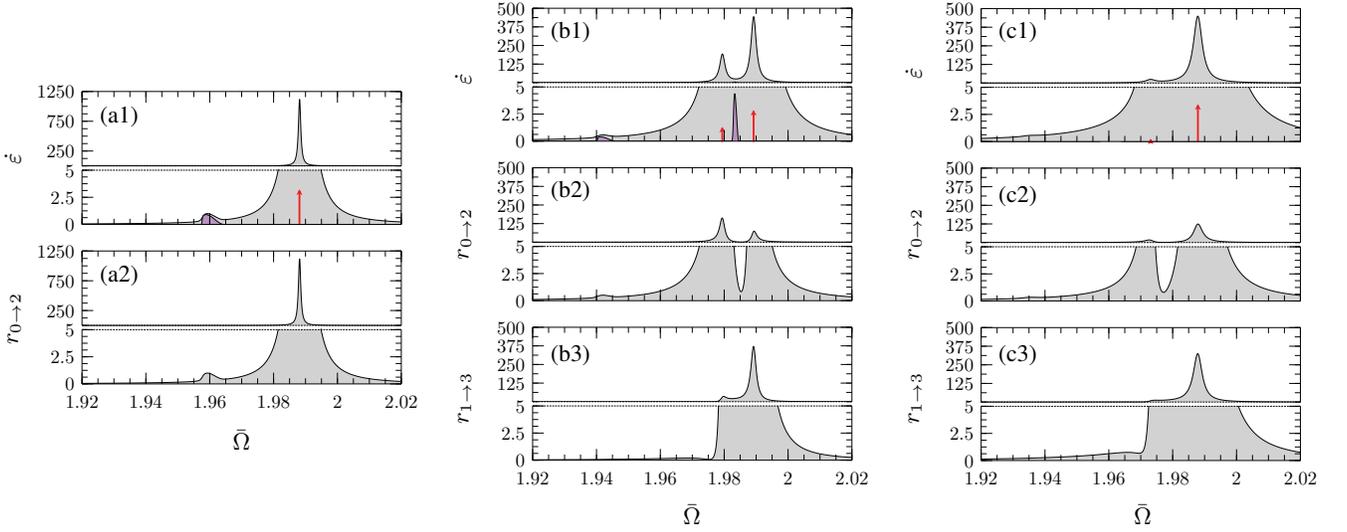

\begin{lpic}[l(0mm),r(-8mm),t(0mm),b(5mm)]{figs/rate_plots_low_T_gamma=0.01(18cm,)}
\lbl[bl]{-3,62,90;$\dot{\varepsilon}$}
\lbl[bl]{-3,28,90;$r_{0 \rightarrow 2}$}
\lbl[bl]{37,8,0;$\bar{\Omega}$}
\lbl[bl]{82,77,90;$\dot{\varepsilon}$}
\lbl[bl]{82,44,90;$r_{0 \rightarrow 2}$}
\lbl[bl]{82,13,90;$r_{1 \rightarrow 3}$}
\lbl[bl]{122,-6,0;$\bar{\Omega}$}
\lbl[bl]{167,78,90;$\dot{\varepsilon}$}
\lbl[bl]{167,44,90;$r_{0 \rightarrow 2}$}
\lbl[bl]{167,13,90;$r_{1 \rightarrow 3}$}
\lbl[bl]{207,-6,0;$\bar{\Omega}$}
\lbl[bl]{12,69,0;(a1)}
\lbl[bl]{12,39,0;(a2)}
\lbl[bl]{97,85,0;(b1)}
\lbl[bl]{97,55,0;(b2)}
\lbl[bl]{97,24,0;(b3)}
\lbl[bl]{181.5,85,0;(c1)}
\lbl[bl]{181.5,54,0;(c2)}
\lbl[bl]{181.5,24,0;(c3)}
\end{lpic}
\caption{Energy absorption rates and inter-subband transition rates for various transverse confinement widths at a fixed dipolar interaction strength $r_d = 0.05$ and temperature $T = 0.1\,T^* \simeq 0.008 \,T_F^{(0)}$. The grey shaded plots correspond to quantities evaluated at a finite pulse switching rate of $\eta = 0.01\,\hbar n / 2m$. The blue plots are evaluated in the adiabatic switching limit and are shown for comparison. The red spikes correspond to excitons lying outside of p-h continuums in the adiabatic limit. The length of the spikes indicate their dimensionless spectral weight, $w$. (a1-2) $\sqrt{n}a_{\perp}=0.30$, (b1-3) $\sqrt{n}a_{\perp}=0.45$, (c1-3) $\sqrt{n}a_{\perp}=0.58$.}
\label{fig:ITRlow}
\end{figure*}

\begin{figure*}[t!]
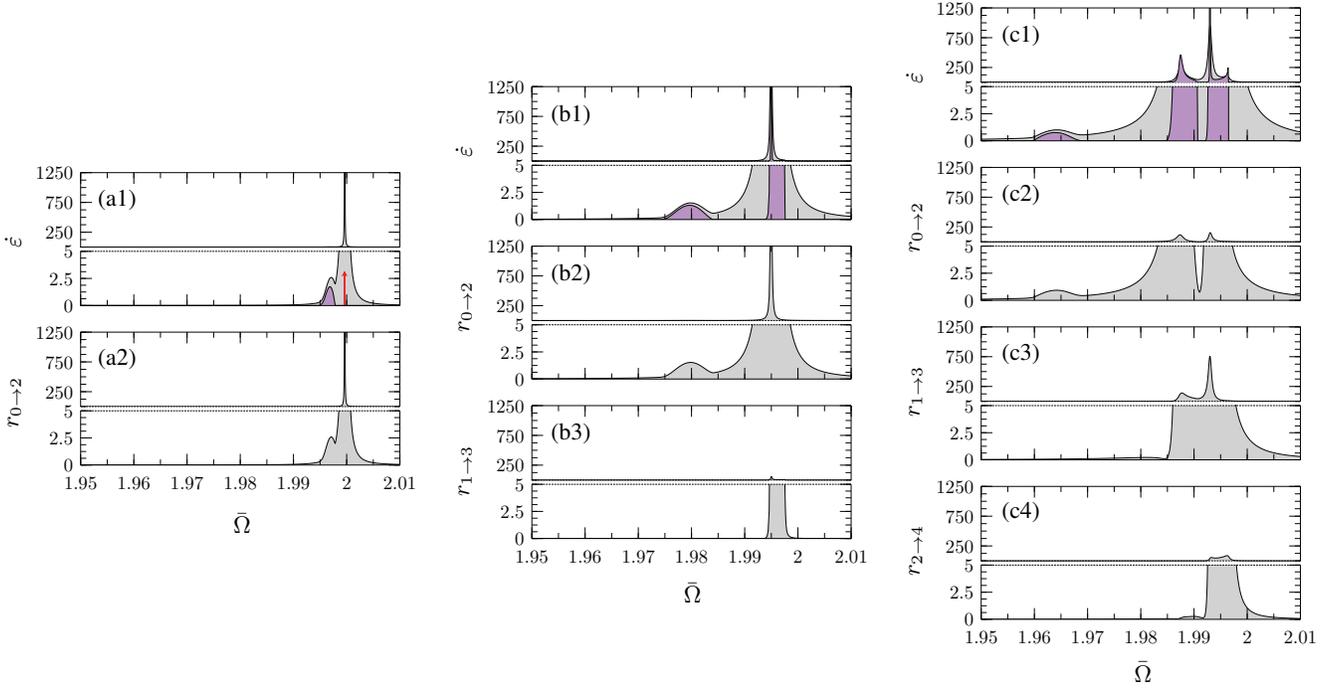

\begin{lpic}[l(0mm),r(0mm),t(0mm),b(5mm)]{figs/rate_plots_high_T_gamma=0.01(17.0cm,)}
\lbl[bl]{-3,76,90;$\dot{\varepsilon}$}
\lbl[bl]{-3,42,90;$r_{0 \rightarrow 2}$}
\lbl[bl]{37,22,0;$\bar{\Omega}$}
\lbl[bl]{82,93,90;$\dot{\varepsilon}$}
\lbl[bl]{82,60,90;$r_{0 \rightarrow 2}$}
\lbl[bl]{82,29,90;$r_{1 \rightarrow 3}$}
\lbl[bl]{122,9,0;$\bar{\Omega}$}
\lbl[bl]{167,107,90;$\dot{\varepsilon}$}
\lbl[bl]{167,75,90;$r_{0 \rightarrow 2}$}
\lbl[bl]{167,44,90;$r_{1 \rightarrow 3}$}
\lbl[bl]{167,14,90;$r_{2 \rightarrow 4}$}
\lbl[bl]{207,-6,0;$\bar{\Omega}$}
\lbl[bl]{11.5,83,0;(a1)}
\lbl[bl]{11.5,53,0;(a2)}
\lbl[bl]{97,99,0;(b1)}
\lbl[bl]{97,69,0;(b2)}
\lbl[bl]{97,39,0;(b3)}
\lbl[bl]{182,114,0;(c1)}
\lbl[bl]{182,84,0;(c2)}
\lbl[bl]{182,54,0;(c3)}
\lbl[bl]{182,24,0;(c4)}
\end{lpic}
\caption{Energy absorption rates and inter-subband transition rates for various transverse confinement widths at a fixed dipolar interaction strength $r_d = 0.05$ and temperature $T = 10\,T^* \simeq 0.8 \,T_F^{(0)}$. The grey shaded plots correspond to quantities evaluated at a finite pulse switching rate of $\eta = 0.01\,\hbar n / 2m$. The blue plots are evaluated in the adiabatic switching limit and are shown for comparison. The red spikes correspond to excitons lying outside of p-h continuums in the adiabatic limit. The length of the spikes indicate their dimensionless spectral weight, $w$. (a1-2) $\sqrt{n}a_{\perp}=0.05$, (b1-3) $\sqrt{n}a_{\perp}=0.13$, (c1-3) $\sqrt{n}a_{\perp}=0.28$.}
\label{fig:ITRhigh}
\end{figure*}
Motivated by band-resolved spectroscopy experiments and also in order to elucidate the nature of excitonic peaks in the energy absorption spectrum, we evaluate the inter-subband transition rates in this section. According to the remarks at the end of Sec.~\ref{sec:TMSmethod} and estimates obtained in Appendix~\ref{app:relax} for the quasiparticle relaxation rate, we choose $\eta = 0.01\,\hbar n/2m$. According to Eq.~(\ref{eq:relaxcrit}), this choice guarantees that the effective duration of the modulation interval is smaller than the lifetime of the majority of the inter-subband excitations.\\

Fig.~\ref{fig:ITRlow} shows the energy absorption rate along with different inter-subband transition rates for three transverse confinement widths at $T = 0.1\,T^* \simeq 0.008 \,T_F^{(0)}$. The topmost plots are slices of Fig.~\ref{fig:EARabot}a, which are broadened due to finite switching-on rate. The energy absorption rates in the adiabatic switching limit are also shown for comparison. Note that we have chosen different scales along the vertical axis for the clarity of presentation.

For $\sqrt{n}a_\perp = 0.30$ (plots a1 and a2), the spectrum consists of a single p-h continuum and an exciton. The only non-vanishing inter-subband transition channel is $0 \rightarrow 2$. For $\sqrt{n}a_\perp = 0.45$ (plots b1, b2 and a3), the spectrum consists of two p-h continuums and two excitons. We find that the excitons have a mixed nature in this case, i.e. although the first and the second excitons are mainly composed of $0 \rightarrow 2$ and $1 \rightarrow 3$ p-h pairs, each has a considerable contribution from the other p-h pairs. For $\sqrt{n}a_\perp = 0.54$ (plots c1, c2 and c3), one finds that the spectral weight of the first exciton is reduced and second exciton dominates. The energy absorption is mainly due to $1 \rightarrow 3$ transitions in this case.\\

Fig.~\ref{fig:ITRhigh} shows the same quantities as Fig.~\ref{fig:ITRlow} at a higher temperature $T = 10\,T^* \simeq 0.8 \,T_F^{(0)}$. For $\sqrt{n}a_\perp = 0.05$ (plots a1 and a2), the spectrum consists of a single p-h continuum and an exciton, both of which are associated to $0 \rightarrow 2$ transitions. For $\sqrt{n}a_\perp = 0.16$ (plots b1, b2 and a3), the spectrum consists of two p-h continuums and one visible exciton which lies inside the second continuum. The exciton is a mixture of both $0 \rightarrow 2$ and $1 \rightarrow 3$ p-h pairs. For $\sqrt{n}a_\perp = 0.28$ (plots c1-c4), one finds more continuums and excitons. Again, the excitons are mixtures of different p-h pairs.\\

Another interesting finding is the significantly reduced spectral weight between the two excitons in the $r_{0\rightarrow 2}$ plots (See Figs.~\ref{fig:ITRlow}-b2,~\ref{fig:ITRlow}-c2 and~\ref{fig:ITRhigh}-c2). This phenomenon is an example of Fano interference between the excitons and has also been previously reported in semiconductor quantum wells~\cite{Nikonov1999}.

\section{Experimental Outlook}\label{sec:disc}
Like numerous other theoretical predictions about many-body phenomena using polar molecules, the observation of the results presented in this study may also be experimentally challenging. In this section, we discuss some of these challenges and point out the experimental signatures that are expected to more robust.\\

The group at JILA have successfully realized a near-degenerate gas of KRb polar molecules with a density of $n \sim 4 \times 10^{11}\,\mathrm{cm}^{-3}$, a temperature of $T \simeq 1.4\,T_F^{(0)}$ and a maximum dipole moment of $D \simeq 0.22~\mathrm{Debye}$~\cite{Ospelkaus2008,Ni2008,Ni2009,Ospelkaus2010,Ni2010}. Loading the gas into a one-dimensional optical lattice with a wavelength of $\sim 1\,\mu m$ yields stacks of pancakes with $\sqrt{n}a_{\perp} \simeq 0.2$ and $r_d \simeq 0.04$, closely matching the parameters used to obtain the higher temperature plots in Fig.~\ref{fig:EARtemp}.

In the experiments with KRb and generally all species with energetically favorable two-body chemical reactions, the molecules have a finite lifetime due to head-to-tail collisions. Populating higher subbands generally results in higher reactive collision rates, making experiments more challenging. From this perspective, it is favorable to prepare the system in the single subband regime. The results shown in Fig.~\ref{fig:EARdip2} belong to such regime and in particular,  observation of the the excitonic peak is expected to be feasible. However, presence of heating noise and broadening of excitonic peaks due to inelastic scatterings and finite pulse durations may mask the p-h continuum, which has less than $1\%$ of the spectral weight. If only a single peak is resolved, it can be difficult to judge if the peak is associated to the exciton. Therefore, one may need to look for signatures of excitons in the multi-subband regime.\\

Qu\'em\'ener {\em et al.} have recently studied chemical reaction rates of reactive polar molecules in pancake geometries and have established that the suppression of chemical reactions remains effective even if the first few excited subbands are populated~\cite{Quemener2011}. Realizing ultracold gases of non-reactive polar molecules is another possibility for making quasi-two-dimensional systems with long lifetimes.

In the multi-subband regime, a more robust experimental signature for excitonic effects can be found in the plots shown in Fig.~\ref{fig:EARabot}. Appearance of two excitonic peaks (one mainly associated to $0 \rightarrow 2$ and the other to $1 \rightarrow 3$) and disappearance of the first exciton upon relaxing the trap is expected to be detectable in both energy-resolved and band-resolved lattice modulation spectroscopy experiments. Most strikingly, the Fano interference effects lowers the spectral weight between the two excitons significantly (see Fig.~\ref{fig:ITRhigh}-c2) and allows the double peak structure to be resolved even if the broadening of peaks is beyond their frequency separation. The plots in Fig.~\ref{fig:EARabot}b and Fig.~\ref{fig:ITRhigh} suggest that observation of excitonic effects is not limited to strongly degenerate systems and is expected to be feasible for temperatures of the order of the Fermi temperature as well.

We also note that extension of the band-resolved measurement technique to momentum-resolved measurements~\cite{Winkler2006,Ernst2010} can be used to directly measure the exciton wavefunctions.\\

Although the band-resolved measurements yield valuable information about the nature of excitations, the energy-resolved measurements are expected to be less prone to broadening since they are not affected by quasiparticle relaxation processes. Therefore, energy-resolved measurements are expected to yield shaper peaks in the spectrum.

We also note that we chose the exponentially switched-on pulse shape in this study mainly for theoretical convenience. Generally, the broadening of the peaks can be minimized by ramping up the amplitude of modulations as fast as possible, followed by a hold interval at the maximum amplitude, rather than a gradual amplitude build-up.

\section{Conclusion}\label{sec:conc}
In this paper, we theoretically analyzed a quasi-two-dimensional system of fermionic polar molecules in a harmonic transverse confining potential. The electric dipole moments of the molecules were assumed to be aligned perpendicular to the confining plane by applying a strong d.c. electric field. We studied the renormalization of the energy bands in the Hartree-Fock approximation for various trap strengths, dipolar interaction strengths and temperatures (Fig.~\ref{fig:band}). The renormalized subbands were found to be parity-conserving mixtures of the non-interacting subbands~(Fig.~\ref{fig:dens}). A phase diagram was obtained for the normal liquid phases of the system at zero temperature~(Fig.~\ref{fig:phasediag}) as a function of trap and interaction strengths.

We also studied the inter-subband excitation spectrum of the system in the conserving time-dependent Hartree-Fock (TDHF) approximation for various system parameters and presented theoretical predictions for energy absorption rates in lattice modulation spectroscopy experiments. We found that the excitation spectrum consists of both inter-subband p-h continuums and anti-bound exciton~(Figs.~\ref{fig:EARdip2}, \ref{fig:EARtemp} and \ref{fig:EARabot}). The excitons capture more than $99\%$ of the spectral weight. Our results indicate that both types of excitations are present for interaction strengths and temperatures accessible in current experiments with polar molecules.

By evaluating the inter-subband transition rates in lattice modulation spectroscopy experiments, we studied the nature of excitons and found that they are generally mixtures of p-h excitations arising from several subbands (Figs.~\ref{fig:ITRlow} and \ref{fig:ITRhigh}). We also found a criteria for the validity of the predictions of time-dependent Hartree-Fock approximation for inter-subband transitions (Eq.~\ref{eq:relaxcrit}).

Finally, we briefly discussed the experimental outlook of this study and pointed out the most robust experimental signatures of excitonic effects based on the the presented results.

\section{Acknowledgements}
The authors would like to thank Bertrand Halperin, Jun Ye, Deborah Jin,   Brian C. Sawyer and David Pekker for insightful discussions. This work was supported by the Army Research Office with funding from the DARPA OLE program, Harvard-MIT CUA, NSF Grant No. DMR-07-05472, AFOSR Quantum Simulation MURI, AFOSR MURI on Ultracold Molecules and the ARO-MURI on Atomtronics.

\appendix

\section{The generating function for the effective inter-subband dipolar interactions in harmonic traps}\label{sec:Vformula}
In this appendix, we derive a generating function for the effective inter-subband dipolar interactions for harmonic traps and provide explicit formulas for the first few. We start the derivation by transforming the in-plane coordinates of the electric dipole-dipole interaction (Eq.~\ref{eq:vdip}) to the momentum space:
\begin{eqnarray}\label{eq:vdipz}
\mathcal{V}_{\mathrm{dip}}(z,\qq) & = & \int \mathrm{d}^2\xx\,e^{-i\mathbf{q}\cdot\xx}\,V_{\mathrm{dip.}}(z,\xx)\nonumber\\
&=& \frac{8 \pi D^2}{3}\delta(z) - 2\pi D^2 |\qq|\,e^{-|\qq||z|}.
\end{eqnarray}
The effective inter-subband interaction is obtained by integrating out the $z$ coordinate:
\begin{widetext}
\begin{eqnarray}\label{eq:vdipzint}
\mathcal{V}_{\alpha\beta;\gamma\lambda}(\qq) &=& \int \mathrm{d}z\,\mathrm{d}z'\,\phi_{\alpha}(z)\,\phi_{\beta}(z)\,\phi_{\gamma}(z')\,\phi_{\lambda}(z')\,\mathcal{V}_{\mathrm{dip}}(z-z',\qq)\nonumber\\
&=&\frac{8\pi D^2}{3} \int_{-\infty}^{\infty} \mathrm{d}z\, \phi_{\alpha}(z)\,\phi_{\beta}(z) \,\phi_{\gamma}(z')\,\phi_{\lambda}(z') - 4 \pi D^2 |\qq|\,\int_0^{\infty} \mathrm{d}\xi\,\int_{-\infty}^{\infty}\mathrm{d}\eta\,\Bigg[\phi_{\alpha}(\eta+\xi)\,\phi_{\beta}(\eta+\xi)\nonumber\\
&&\times\,\phi_{\gamma}(\eta-\xi)\,\phi_{\lambda}(\eta-\xi)+\phi_{\alpha}(\eta-\xi)\,\phi_{\beta}(\eta-\xi)\,\phi_{\gamma}(\eta+\xi)\,\phi_{\lambda}(\eta+\xi)\Bigg]\, e^{-2|\qq|\xi},
\end{eqnarray}
\end{widetext}
where the transverse wavefunctions, $\phi_{\alpha}(z)$, etc., are the well-known harmonic oscillator wavefunctions (Eq.~\ref{eq:ensubband}). We have changed variables to $\eta = (z+z')/2$ and $\xi = (z-z')/2$ in order to take care of the absolute value appearing in Eq.~(\ref{eq:vdipz}) conveniently.

A generating function for $\mathcal{V}_{\alpha\beta;\gamma\lambda}(\qq)$ can be found by expressing the Hermite functions appearing in Eq.~(\ref{eq:ensubband}) in terms of their generating function:
\begin{equation}\label{eq:hermite}
e^{2xt-t^2} = \sum_{n=0}^{\infty}\frac{H_n(x)\,t^n}{n!}.
\end{equation}
Plugging Eq.~(\ref{eq:ensubband}) into Eq.~(\ref{eq:vdipzint}) and using Eq.~(\ref{eq:hermite}), we get the following generating function for $\mathcal{V}_{\alpha\beta;\gamma\lambda}(\qq)$:
\begin{widetext}
\begin{eqnarray}\label{eq:gen}
\Gamma(w_1,w_2,w_3,w_4;\qq) &=&\frac{8 D^2}{3 a_{\perp}} \int_{-\infty}^{\infty} \mathrm{d}\tilde{z}\, e^{2\tilde{z}w_1-w_1^2}\,e^{2\tilde{z}w_2-w_2^2} \,e^{2\tilde{z}w_3-w_3^2} \,e^{2\tilde{z}w_4-w_4^2}\,e^{-4\tilde{z}^2/2}\nonumber\\
&&- 4 D^2 |\qq|\,\int_0^{\infty} \mathrm{d}\xi\,\int_{-\infty}^{\infty}\mathrm{d}\eta\,\Bigg[e^{2(\eta+\xi)w_1-w_1^2}\,e^{2(\eta+\xi)w_2-w_2^2}\,e^{2(\eta-\xi)w_3-w_3^2}\,e^{2(\eta-\xi)w_4-w_4^2}\,e^{-(\eta+\xi)^2}e^{-(\eta-\xi)^2}\nonumber\\
&&+\,e^{2(\eta-\xi)w_1-w_1^2}\,e^{2(\eta-\xi)w_2-w_2^2}\,e^{2(\eta+\xi)w_3-w_3^2}\,e^{2(\eta+\xi)w_4-w_4^2}\,e^{-(\eta+\xi)^2}\,e^{-(\eta-\xi)^2}\Bigg]\, e^{-2|\qq|\xi}\nonumber\\
& = & \frac{4\sqrt{2\pi}D^2}{3 a_{\perp}}\,e^{-w_1^2-w_2^2-w_3^2-w_4^2+\frac{1}{2}(w_1+w_2+w_3+w_4)^2}\nonumber\\
&&- \pi D^2 |\qq|\,e^{|\qq|^2 a_{\perp}^2/2 + 2 w_1 w_2 + 2 w_3 w_4 + |\qq|a_{\perp}(w_1+w_2-w_3-w_4)}\,\mathrm{Erfc}\left(\frac{|\qq|a_{\perp} + w_1 + w_2 - w_3 - w_4}{\sqrt{2}}\right)\nonumber\\
&&- \pi D^2 |\qq|\,e^{|\qq|^2 a_{\perp}^2/2 + 2 w_3 w_4 + 2 w_1 w_2 + |\qq|a_{\perp}(w_3+w_4-w_1-w_2)}\,\mathrm{Erfc}\left(\frac{|\qq|a_{\perp} + w_3 + w_4 - w_1 - w_2}{\sqrt{2}}\right),
\end{eqnarray}
\end{widetext}
and $\mathcal{V}_{\alpha\beta;\gamma\lambda}(\qq)$ is given by:
\begin{equation}\label{eq:genfunc}
\mathcal{V}_{\alpha\beta;\gamma\lambda}(\qq) = \frac{\partial_{w_1}^{\alpha}\partial_{w_2}^{\beta}\partial_{w_3}^{\gamma}\partial_{w_4}^{\lambda}\,\Gamma(w_1,w_2,w_3,w_4;\qq)}{\sqrt{\alpha!\,\beta!\,\gamma!\,\lambda!\,2^{\alpha}\,2^{\beta}\,2^{\gamma}\,2^{\lambda}}}\bigg|_{w_i=0}.
\end{equation}
Using Eq.~(\ref{eq:genfunc}), we provide explicit expressions for the interactions between the first two subbands and their long wavelength limits for reference:
\begin{eqnarray}
\mathcal{V}_{00;00}(\qq) &=& \frac{4\sqrt{2\pi}D^2}{3 a_{\perp}} - 2 \pi D^2 |\qq| e^{|\qq|^2 a_{\perp}^2/2}\nonumber\\
&&\times\,\mathrm{Erfc}(|\qq|a_{\perp}/\sqrt{2})\nonumber\\
&\simeq&\frac{4\sqrt{2\pi}D^2}{3 a_{\perp}} - 2\pi D^2 |\qq| + \mathcal{O}(|\qq|^2),
\end{eqnarray}
\begin{eqnarray}
\mathcal{V}_{01;01}(\qq) &=& \frac{\sqrt{2\pi}D^2}{3 a_{\perp}}(2 - 3 |\qq|^2 a_{\perp}^2) + \pi D^2 |\qq|^3 a_\perp^2 e^{|\qq|^2 a_{\perp}^2/2}\nonumber\\
&&\times\,\mathrm{Erfc}(|\qq|a_{\perp}/\sqrt{2})\nonumber\\
&\simeq& \frac{2\sqrt{2\pi}D^2}{3 a_{\perp}} - \sqrt{2\pi} D^2 |\qq|^2 a_{\perp} + \mathcal{O}(|\qq|^3),
\end{eqnarray}
\begin{eqnarray}
\mathcal{V}_{00;11}(\qq) &=& \frac{\sqrt{2\pi}D^2}{3 a_{\perp}}(2 + 3 |\qq|^2 a_{\perp}^2) - \pi D^2 |\qq| (2+|\qq|^2 a_\perp^2)\nonumber\\
&&\times\,e^{|\qq|^2 a_{\perp}^2/2}\,\mathrm{Erfc}(|\qq|a_{\perp}/\sqrt{2})\nonumber\\
&\simeq& \frac{2\sqrt{2\pi}D^2}{3 a_{\perp}} - 2\pi D^2 |\qq| + \mathcal{O}(|\qq|^2),
\end{eqnarray}
\begin{eqnarray}
\mathcal{V}_{11;11}(\qq) &=& \frac{\sqrt{2\pi}D^2}{2 a_{\perp}}(1 + |\qq|^2 a_{\perp}^2)(2 + |\qq|^2 a_{\perp}^2)\nonumber\\
&&- \frac{\pi D^2}{2} |\qq| (2+|\qq|^2 a_\perp^2)^2\,e^{|\qq|^2 a_{\perp}^2/2}\nonumber\\
&&\times\,\mathrm{Erfc}(|\qq|a_{\perp}/\sqrt{2})\nonumber\\
&\simeq& \frac{\sqrt{2\pi}D^2}{a_{\perp}} - 2\pi D^2 |\qq| + \mathcal{O}(|\qq|^2).
\end{eqnarray}
Note that $\mathcal{V}_{01;01} = \mathcal{V}_{10;01} = \mathcal{V}_{10;10} = \mathcal{V}_{01;10}$, $\mathcal{V}_{00;11} = \mathcal{V}_{11;00}$ and $\mathcal{V}_{00;01} = \mathcal{V}_{00;10} = \mathcal{V}_{01;00} = \mathcal{V}_{10;00} = 0$.

\section{Numerical solution of the Hartree-Fock equation}\label{sec:HFnum}
Our goal is to develop a numerical routine to find a self-consistent solution to Eq.~(\ref{eq:finalhf}) at any given temperature, dipolar interaction strength and transverse confinement strength. In its current form, Eq.~(\ref{eq:finalhf}) describes an infinite number of coupled non-linear integral equations. However, we note that at zero temperature, there is only a finite number of occupied subbands due to the sharp step-like behavior of Fermi occupation function. At finite temperatures, we also expect to find a finite number of subbands with a significant population. Thus, a cut-off can be imposed on the number of subbands in practice. In our implementation, we choose the cut-off such that the fractional density of the highest neglected subband is less than $10^{-3}$. This ensures that the presence of higher subbands have a negligible effect, i.e. of the order of $10^{-3} D^2\,k_F^3$, on the energies of the lower subbands (See Eq.\ref{eq:finalhf}). The number of dimensions of the integral equations can also be reduced due to the isotropy of the inter-subband interactions, which is inherited by the self-energy matrices and their related quantities. The self-consistent Hartree-Fock self-energy equation can be rewritten as:
\begin{eqnarray}\label{eq:finalhf2}
\Sigma^{\star}_{\mu\nu}(k) &=& \int \frac{k'\,\mathrm{d}k'}{2\pi}\left[\mathcal{W}_{\mu\nu;\gamma\lambda}(0,0)-\mathcal{W}_{\mu\lambda;\gamma\nu}(k,k')\right]U(k')_{\lambda\rho}\nonumber\\&&\times\,U(k')_{\gamma\rho}\,n^F\left[\xi_{\rho}(k')\right],
\end{eqnarray}
where $\mathcal{W}_{\mu\lambda;\gamma\nu}(k,k')$ is the angle-averaged interaction, defined as:
\begin{equation}
\mathcal{W}_{\mu\lambda;\gamma\nu}(k,k') = \int_0^{2\pi}\frac{\mathrm{d}\phi}{2\pi}\, \mathcal{V}_{\mu\lambda;\gamma\nu}\left(\sqrt{q^2 + k^2 - 2qk\cos\phi}\right).
\end{equation}
From a numerical perspective, it is favorable to deal with integrals with bounded integration domains. Although the momentum integral appearing Eq.~(\ref{eq:finalhf2}) is unbounded, an upper bound can be imposed on it in a controlled way. Since the subbands have (approximately) a quadratic energy dispersion for large momenta, the Fermi occupation of states falls super-exponentially fast for large $k'$ and the integral kernel of Eq.~(\ref{eq:finalhf2}) becomes negligible for large $k'$. In our implementation, we impose a large momentum cut-off $K_{\mathrm{cut}}$, such that $n^F\left[\xi_{\rho}(k')\right]<10^{-6}$ for all $k'\geq K_{\mathrm{cut}}$. Finally, we approximate the resulting bounded integrations using quadrature formulas:
\begin{eqnarray}\label{eq:quad}
&&\hspace{-10pt}\Sigma^{\star}(k_i)_{\mu\nu} = \sum_{j=1}^{N_p} \frac{w_j k'_j}{2\pi}\left[\mathcal{V}_{\mu\nu;\gamma\lambda}(0)-\mathcal{W}_{\mu\lambda;\gamma\nu}(k_i,k'_j)\right]\nonumber\\&&\,\,\,\times\,U(k'_j)_{\lambda\rho}U(k'_j)_{\gamma\rho}n^F\left[\xi_{\rho}(k'_j)\right], \quad i = 1, \ldots, N_p
\end{eqnarray}
where $k_i$ and $w_i$ denote the quadrature nodes and weights respectively and $N_p$ is the number of quadrature points. We used a 200-point Gauss-Lobatto quadrature in our implementation~\cite{Abramowitz}.\\

It is desirable from an experimental perspective to solve the Hartree-Fock equations for a given particle density $n$ instead of the chemical potential $\mu$. Therefore, the chemical potential must be found in a such a way that the self-energy equation (Eq.~\ref{eq:quad}) and particle density equation,
\begin{eqnarray}\label{eq:Npart}
n &=& \sum_{\rho}\int \frac{\mathrm{d}^2\mathbf{k}}{(2\pi)^2}\,n^F\left[\xi_{\rho}(\mathbf{k})\right]\simeq \sum_{\rho}\sum_{j=1}^{N_p} \frac{w_j k_j}{2\pi}n^F\left[\xi_{\rho}(k_j)\right],\nonumber\\
\end{eqnarray}
are satisfied simultaneously. We found it more efficient to solve the system of non-linear equations (Eqs.~\ref{eq:quad} and~\ref{eq:Npart}) using an interior reflective trust region method~\cite{Coleman1996} instead of the usual solution by iterations. In all of the runs, the solution was unique and the convergence was rapid.

\section{Energy absorption rate for an exponentially switched-on a.c. modulation pulse}\label{app:Edot}
In this appendix, we derive the formula given for the energy absorption rate for an exponentially switched-on a.c. modulation pulse in Sec.~\ref{sec:TMS} (Eq.~\ref{eq:Edot}). Our strategy is to evaluate the total absorbed energy for a given pulse shape, $\Delta E$. We define the energy absorption rate as $\Delta E / \Delta t$, where $\Delta t$ is the effective time interval during which the external field interacts with the system. For the exponentially switched on and off pulse of Eq.~(\ref{eq:pulse}), $\Delta t \sim 2 \eta^{-1}$. Our starting point is Eq.~(\ref{eq:Edot0}), which we integrate to get $\Delta E$:
\begin{align}\label{eq:Edot2}
\Delta E & = i\xi^2\,\int_{-\infty}^{\infty}\mathrm{d}t\,f(t) \int_{-\infty}^{\infty}\mathrm{d}t'\,f(t')\,\partial_t\,\mathcal{T}(t-t')\nonumber\\
&= i\xi^2 \int \frac{\mathrm{d}\omega}{2\pi}\,|f(\omega)|^2 \, \omega\,\mathcal{T}(\omega).
\end{align}
Here, $f(\omega)$ is the Fourier transform of Eq.~(\ref{eq:pulse}). We write $|f(\omega)|^2$ as:
\begin{equation}
|f(\omega)|^2 = f_+^2(\omega)+f_-^2(\omega) + 2 f_+(\omega)f_-(\omega),
\end{equation}
where:
\begin{equation}
f_\pm(\omega) = \frac{\eta}{\eta^2 + (\omega\mp\Omega)^2}.
\end{equation}

First, we focus on the contributions of $f_\pm^2(\omega)$ to $\Delta E$. This function has two second order poles at frequencies $\mp \Omega + i\eta$ and $\pm \Omega - i\eta$. Since $\mathcal{T}(t)$ is a causal function, $\mathcal{T}(\omega)$ is analytic in the upper complex plane and the integral in the second line of Eq.~(\ref{eq:Edot2}) can be evaluated by closing the contour on the upper half-plane. The resulting contribution is easily found to be:
\begin{multline}
\Delta E_\pm \equiv -\xi^2 \eta^2 \Bigg[\frac{\mathcal{T}(\pm \Omega + i\eta)}{(2i\eta)^2}+\frac{(\pm \Omega + i\eta)\mathcal{T}'(\pm \Omega + i\eta)}{(2i\eta)^2}\\
-\frac{2(\pm\Omega+i\eta)\mathcal{T}(\pm \Omega + i\eta)}{(2i\eta)^3}\Bigg],
\end{multline}
where $\mathcal{T}'(\omega) = (\mathrm{d}/\mathrm{d}\omega)\mathcal{T}(\omega)$.\\

The contribution of the cross-term, $2 f_+(\omega)f_-(\omega)$, to $\Delta E$ can be evaluated in the same way and we get:
\begin{equation}
\Delta E_c \equiv \frac{i\xi^2\eta}{8\Omega} \big[\mathcal{T}(\Omega + i\eta)-\mathcal{T}(-\Omega + i\eta)\big].
\end{equation}

Using Eqs.~(\ref{eq:defT}),~(\ref{eq:defT1}),~(\ref{eq:integPI}) and~(\ref{eq:barepol}), it is straightforward to establish the following identities:
\begin{align}
\mathcal{T}(\omega)&=\mathcal{T}(-\omega),\nonumber\\
\mathcal{T}(\omega^*)&=\mathcal{T}(\omega)^*,\nonumber\\ 
\mathcal{T}'(\omega)&=-\mathcal{T}'(-\omega),\nonumber\\
\mathcal{T}'(\omega^*)&=\mathcal{T}'(\omega)^*,
\end{align}
using which we get $\mathcal{T}(-\Omega+i\eta)=\mathcal{T}(\Omega+i\eta)^*$ and $\mathcal{T}'(-\Omega+i\eta)=-\mathcal{T}'(\Omega+i\eta)^*$. Using these identities, we get the following simple expression for $\dot{E} = \eta\,\Delta E / 2 \equiv \eta\,(\Delta E_+ + \Delta E_- + \Delta E_c)/2$:
\begin{multline}
\dot{E} = \frac{\eta}{4}\,\mathfrak{Re}\left[\mathcal{T}(\Omega+i\eta)+(\Omega+i\eta)\mathcal{T}'(\Omega+i\eta)\right]\\
-\frac{1}{4}\,\mathfrak{Im}\left[(\Omega+i\eta)\mathcal{T}(\Omega+i\eta)\right].
\end{multline}
In the limit $\eta / \Omega \ll 1$, the term in the second line is dominant and we get the desired result (Eq.~\ref{eq:Edot}).

\section{Estimation of the relaxation rate of inter-subband p-h excitations}\label{app:relax}
\begin{figure}[t]
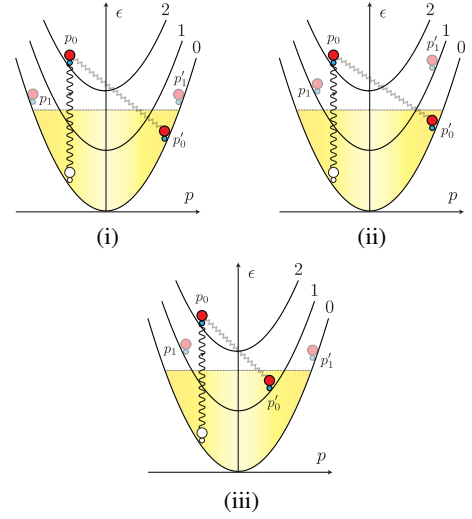

\begin{lpic}[l(0mm),r(0mm),t(0mm),b(5mm)]{figs/relaxation(6cm,)}
\lbl[bl]{26,76,0;(i)}
\lbl[bl]{108,76,0;(ii)}
\lbl[bl]{66,-7,0;(iii)}
\end{lpic}
\caption{The three relaxation processes contributing to $\Gamma^{\mathrm{rel.}}_2$.}
\label{fig:relax}
\end{figure}
\begin{figure*}[t!]
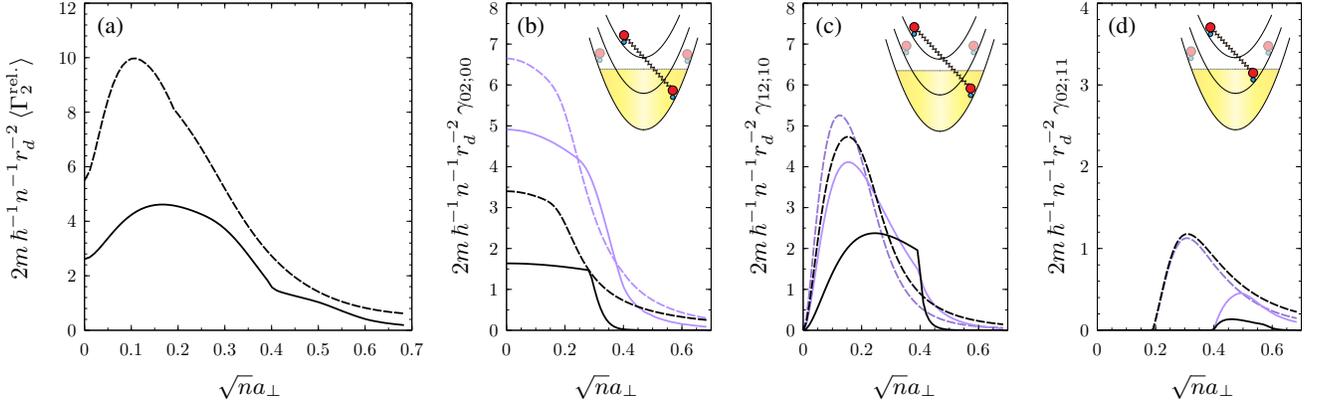

\begin{lpic}[l(4mm),r(0mm),t(0mm),b(3mm)]{figs/relaxation_plots(17cm,)}
\lbl[bl]{0,40,90;$2m\,\hbar^{-1} n^{-1} r_d^{-2}\,\langle\Gamma^{\mathrm{rel.}}_2\rangle$}
\lbl[bl]{173,42,90;$2m\,\hbar^{-1} n^{-1} r_d^{-2}\,\gamma_{02;00}$}
\lbl[bl]{290,42,90;$2m\,\hbar^{-1} n^{-1} r_d^{-2}\,\gamma_{12;10}$}
\lbl[bl]{407,42,90;$2m\,\hbar^{-1} n^{-1} r_d^{-2}\,\gamma_{02;11}$}
\lbl[bl]{72,-7,0;$\sqrt{n}a_\perp$}
\lbl[bl]{212,-7,0;$\sqrt{n}a_\perp$}
\lbl[bl]{329,-7,0;$\sqrt{n}a_\perp$}
\lbl[bl]{446,-7,0;$\sqrt{n}a_\perp$}
\lbl[b]{30,135,0;(a)}
\lbl[b]{195,135,0;(b)}
\lbl[b]{312,135,0;(c)}
\lbl[b]{427,135,0;(d)}
\end{lpic}
\caption{(a) the total momentum-averaged relaxation rate of a quasiparticle in the second excited subband, $\langle\Gamma^{\mathrm{rel.}}_2\rangle$ as function of transverse confinement width $a_\perp$. The solid and dashed lines correspond to $T=0$ and $T=10\,T^*$ respectively. (b), (c) and (d) show the momentum-resolved contribution of different relaxation processes. The black and blue plots correspond to $|\pp_0|=0$ and $|\pp_0|=p_{F,0}$ respectively, where $p_{F,0} \equiv \hbar \sqrt{4\pi n}$, is the Fermi momentum of the zeroth subband at zero temperature. The inset figures are graphical representations of the processes.}
\label{fig:relax}
\end{figure*}
In this appendix, we estimate the relaxation rate of inter-subband p-h excitations created by the lattice modulation pulse. As mentioned in the final remarks of Sec.~\ref{sec:TMSmethod}, such relaxation processes are absent in the TDHF approximation. Therefore, predictions of TDHF approximation for the inter-subband transition rates are only valid if the lifetime of excitations are larger than the pulse duration (See Eq.~\ref{eq:relaxcrit}). 

For weak interactions, we expect the relaxation rate of both types of inter-subband excitations, quasiparticle-like excitations and excitons, to be of the same order. Therefore, we study only the former case here. We are interested in relaxation processes that change the population of subbands and therefore, we neglect the intra-subband scatterings.\\

Relaxation of a quasiparticle excitation can occur through several channels. The number of relaxation channels is higher for quasiparticles in higher subbands. As mentioned in Sec.~\ref{sec:mod}, the inter-subband interaction matrix elements conserve the net parity of the interacting particles and this condition constrains the number of relaxation channels. A simple combinatorial analysis shows that the number of relaxation channels scales like $\sim~N^3 / 4$, where $N$ is the subband index of the decaying quasiparticle. One can, however, argue that the most important contributions still result from the processes involving only the first few lower subbands. The reason is twofold: (1) the inter-subband interaction between particles in well separated subbands is small, and (2) the lower subbands have larger Fermi momenta and consequently, creating a quasiparticle excitation in them upon de-exciting the original quasiparticle has a higher energy cost. Therefore, the quasiparticle relaxation rate of particles in the higher subbands are expected to be of the same order as those in the lower subbands.\\

As a concrete example, we study the relaxation of a quasiparticle excitation in the second excited subband, as a function of transverse confinement width and temperature. Fig.~\ref{fig:relax} shows the three relaxation processes that that contribute to the decay. Since the interactions are assumed to be weak ($r_d \ll 1$), Born's approximation is applicable. We ignore the self-energy correction since such contributions are beyond Born's approximation. The initial and final wavefunctions of the system in all of the processes shown in Fig.~\ref{fig:relax} can be written as $|i\rangle = \tilde{c}^{\dagger}_{\mathbf{p}_0,\lambda}\tilde{c}^{\phantom{\dagger}}_{\mathbf{p}_0,\alpha}|\Psi_0\rangle$ and $|f\rangle = \tilde{c}^{\dagger}_{\mathbf{p}_1,\gamma}\tilde{c}^{\dagger}_{\mathbf{p}'_1,\mu}\tilde{c}^{\phantom{\dagger}}_{\mathbf{p}'_0,\nu}\tilde{c}^{\phantom{\dagger}}_{\mathbf{p}_0,\alpha}|\Psi_0\rangle$ respectively, where $|\Psi_0\rangle$ is the Fermi gas-like ground state of the system. The rate of this generic process can be obtained using Fermi's golden rule: 
\begin{align}
&\Gamma_{\gamma\lambda;\mu\nu}(\mathbf{p}_0,\mathbf{p}'_0,\mathbf{p}_1,\mathbf{p}'_1) =\nonumber\\
&\hspace{30pt}\frac{8\pi^3}{A^3} \left|\tilde{\mathcal{V}}_{\gamma\lambda;\mu\nu}(\pp_1-\pp_0)-\tilde{\mathcal{V}}_{\mu\lambda;\gamma\nu}(\pp'_1-\pp_0)\right|^2\nonumber\\
&\hspace{30pt}\times\,\delta(\pp_0+\pp'_0-\pp_1-\pp'_1)\nonumber\\
&\hspace{30pt}\times\,\delta(\tilde{\xi}_{\lambda}(\pp_0) + \tilde{\xi}_{\nu}(\pp'_0) - \tilde{\xi}_{\gamma}(\pp_1) - \tilde{\xi}_{\mu}(\pp'_1))\nonumber\\
&\hspace{30pt}\times\,\left[1-n_{\gamma}(\pp_1)\right]\,\left[1-n_{\mu}(\pp'_1)\right]\,n_{\nu}(\pp'_0).
\end{align}
We work in units in which $\hbar=1$ in this appendix. Since the particles are identical, $\Gamma_{\gamma\lambda;\mu\nu}=\Gamma_{\mu\lambda;\gamma\nu}$ and over-counting must be carefully avoided. The total contribution of this process can be found by summing over $\pp'_0$, $\pp_1$ and $\pp'_1$. After a lengthy but straightforward algebra, we obtain:
\begin{align}\label{eq:triple}
\Gamma_{\gamma\lambda;\mu\nu}(\mathbf{p}_0) &\equiv A^3\int \frac{\mathrm{d}^2\pp'_0}{(2\pi)^2}\,\frac{\mathrm{d}^2\pp_1}{(2\pi)^2}
\,\frac{\mathrm{d}^2\pp'_1}{(2\pi)^2}\nonumber\\
&\hspace{80pt}\gamma_{\gamma\lambda;\mu\nu}(\mathbf{p}_0,\mathbf{p}'_0,\mathbf{p}_1,\mathbf{p}'_1)\nonumber\\
&=\frac{m}{8\pi^3}\int_0^{\infty}\,Q\,\mathrm{d}Q\int_0^{2\pi}\mathrm{d}\phi\int_0^{2\pi}\mathrm{d}\psi\nonumber\\
&\hspace{15pt}\bigg|\mathcal{V}_{\gamma\lambda;\mu\nu}\left(\frac{1}{2}[\QQ+\qq_0-\pp_0]\right)\nonumber\\
&\hspace{50pt}-\mathcal{V}_{\mu\lambda;\gamma\nu}\left(\frac{1}{2}[\QQ-\qq_0-\pp_0]\right)\bigg|^2\nonumber\\
&\hspace{15pt}\times n_\nu(\QQ)\,\left[1-n_{\gamma}\left(\frac{1}{2}[\QQ+\qq_0+\pp_0]\right)\right]\nonumber\\
&\hspace{15pt}\times\left[1-n_{\mu}\left(\frac{1}{2}[\QQ-\qq_0+\pp_0]\right)\right],
\end{align}
where $\QQ = Q (\cos \phi\,\hat{x}+\sin \phi\,\hat{y})$, $\qq_0= q_0 (\cos \psi\,\hat{x}+\sin \psi\,\hat{y})$ and $q_0=\sqrt{(\lambda+\nu-\gamma-\mu)\,2m\,\hbar\omega_{\mathrm{trap}}+|\QQ-\pp_0|^2}$. The total relaxation rate is obtained by summing over all channels:
\begin{multline}
\Gamma^{\mathrm{rel.}}_{2}(\pp_0) = \Gamma_{02;00}(\pp_0) + \Gamma_{12;10}(\pp_0) + \Gamma_{02;11}(\pp_0).
\end{multline}
We define the momentum-averaged relaxation rate as:
\begin{equation}
\langle \Gamma^{\mathrm{rel.}}_2 \rangle \equiv \frac{\displaystyle\int \frac{\mathrm{d}^2\pp}{(2\pi)^2}\,\left[n_0(\pp)-n_2(\pp)\right]\,\Gamma^{\mathrm{rel.}}_2(\pp)}{\displaystyle\int \frac{\mathrm{d}^2\pp}{(2\pi)^2}\,\left[n_0(\pp)-n_2(\pp)\right]},
\end{equation}
where the averaging is weighted according to the momentum distribution of the p-h excitations generated by the lattice modulation pulse. For weak interactions, the number of excitations at momentum $\pp$ is proportional to the equilibrium occupation of the states, i.e. $\propto n_0(\pp)[1-n_2(\pp)]-n_2(\pp)[1-n_0(\pp)] = n_0(\pp) - n_2(\pp)$.\\
 
We evaluate the triple integral in Eq.~(\ref{eq:triple}) numerically using an adaptive Monte Carlo integration algorithm. Fig.~\ref{fig:relax}a shows $\langle\Gamma^{\mathrm{rel.}}_{2}\rangle$ as a function of transverse confinement width $a_\perp$ for $T=0$ and $T=10\,T^*$. Fig.~\ref{fig:relax}b-d show the momentum-resolved rate of each of the relaxation processes as a function of $a_\perp$ for $|\pp_0| = 0$ and $|\pp_0| = p_{F,0}$, where $p_{F,0} \equiv \hbar \sqrt{4\pi n}$, is the Fermi momentum of the zeroth subband at zero temperature. 

We find that the major contributions result from the first two processes, the former being dominant for stronger traps. Also, increasing the temperature and quasiparticle momenta naturally results in higher relaxation rates. According to Fig.~\ref{fig:relax}a, the highest relaxation rate is $\langle\Gamma^{\mathrm{rel.}}_2\rangle_{\mathrm{max}} \approx 5\,r_d^2\,[\hbar n/2m]$ and $10\,r_d^2\,[\hbar n/2m]$ for $T=0$ and $T=10\,T^*$ respectively. Using these estimates and Eq.~(\ref{eq:relaxcrit}), choosing $\eta \gtrsim 5\,r_d^2\,[\hbar n/2m]$ guarantees that the predictions of TDHF approximation for inter-subband transition rates are reliable for temperatures up to $T = 10\,T^*$. For $r_d = 0.05$, we get $\eta \gtrsim 0.01\,[\hbar n/2m]$.

\section{Numerical Evaluation of $\tilde{h}_{\alpha\beta}(\omega)$}\label{sec:TMSnum}
In this appendix, we discuss the numerical method used to solve Eq.~(\ref{eq:integh}). Our approach is similar to similar to that described in Appendix~\ref{sec:HFnum}, i.e. we approximate the integral equations using integral quadratures. However, given that the integral equation to be solved is linear, the resulting system of equations is also linear.

As a first step, we note that the dimension of the integral equations can be reduced due to the isotropy of the inter-subband interactions and Eq.~(\ref{eq:integh}) reduces to:
\begin{eqnarray}\label{eq:integh1D}
&&\tilde{h}_{\alpha\beta}(k_1;\omega) = \tilde{\mathbf{T}}_{\alpha\beta}(k_1)\nonumber\\
&&\quad-\int\frac{k'\mathrm{d}k'}{2\pi}\,\tilde{\mathcal{W}}_{\alpha\beta;\mu\nu}^{(\mathrm{d}+\mathrm{e})}(k_1,k')\,\tilde{\Pi}^{(0)}_{\mu\nu}(k';\omega)\,\tilde{h}_{\mu\nu}(k';\omega),\nonumber\\
\end{eqnarray}
where:
\begin{eqnarray}
&&\mathcal{\tilde{W}}^{(\mathrm{d+e})}_{\alpha\beta;\mu\nu}(k,k')\nonumber\\
&&\hspace{20pt}= \int_0^{2\pi}\frac{\mathrm{d}\phi_{\kk'}}{2\pi}\,\mathcal{\tilde{U}}^{(\mathrm{d+e})}_{\alpha\beta;\mu\nu}(\kk,\kk')\nonumber\\
&&\hspace{20pt}=\int_0^{2\pi}\frac{\mathrm{d}\phi_{\kk'}}{2\pi}\,\bigg[\mathcal{\tilde{V}}_{\nu\beta;\alpha\mu}(k,k',\sqrt{k^2 + k'^2 - 2kk'\cos\phi_\kk'})\nonumber\\\
&&\hspace{80pt}-\mathcal{\tilde{V}}_{\alpha\beta;\nu\mu}(k,k',0)\bigg].
\end{eqnarray}
First, we study the subtler case of adiabatic switching ($\eta \rightarrow 0$) for which  $\omega = \Omega + i\eta \rightarrow \Omega + i0^+$. In this limit, it is useful to separate the regular and singular parts of $\tilde{\Pi}^{(0)}_{\alpha\beta}(k',\omega)$: 
\begin{multline}\label{eq:bare1}
\tilde{\Pi}^{(0)}_{\alpha\beta}(k',\omega) = P.V.\,\frac{n_{\beta}(k)-n_{\alpha}(k)}{\omega + \tilde{\xi}_{\beta}(k) - \tilde{\xi}_{\alpha}(k)}\\
-i\pi\left[n_{\beta}(k)-n_{\alpha}(k)\right]\delta\big(\Omega + \tilde{\xi}_{\beta}(k) - \tilde{\xi}_{\alpha}(k)\big).
\end{multline}
where $P.V.$ denotes the Cauchy's principal value integration. The delta function may also be written as:
\begin{equation}\label{eq:bare2}
\delta\big(\Omega + \tilde{\xi}_{\beta}(k) - \tilde{\xi}_{\alpha}(k)\big) = \sum_{j}\frac{\delta(k-q_j)}{|\tilde{\xi}_{\alpha}'(q_j)-\tilde{\xi}_{\beta}'(q_j)|},
\end{equation}
where $\{q_j\}$ is the set of solutions of $\Omega + \tilde{\xi}_{\beta}(q) - \tilde{\xi}_{\alpha}(q)=0$. Inserting Eqs.~(\ref{eq:bare1}) and~(\ref{eq:bare2}) into Eq.~(\ref{eq:integh1D}) and approximating the principal value integrals using quadratures, we get:
\begin{eqnarray}\label{eq:quadh}
&&\tilde{h}_{\alpha\beta}(k^{\alpha\beta}_i;\omega) = \delta_{\alpha\gamma}\delta_{\beta\lambda}\tilde{\mathbf{T}}_{\alpha\beta}(k^{\alpha\beta}_i)-\sum_{\mu\nu}\sum_{j=1}^{N_p^{\mu\nu}}\bigg[\frac{w^{\mu\nu}_j\,k^{\mu\nu}_j}{2\pi}\nonumber\\
&&\hspace{10pt}\times\,\tilde{\mathcal{W}}_{\alpha\beta;\mu\nu}^{(\mathrm{d}+\mathrm{e})}(k^{\alpha\beta}_i,k^{\mu\nu}_j)\,\frac{n_{\nu}(k^{\mu\nu}_j)-n_{\mu}(k^{\mu\nu}_j)}{\omega + \tilde{\xi}_{\nu}(k^{\mu\nu}_j) - \tilde{\xi}_{\mu}(k^{\mu\nu}_j)}\nonumber\\
&&\hspace{10pt}\times\,\tilde{h}_{\mu\nu}(k^{\mu\nu}_j;\omega)\bigg]+\,\frac{i}{2}\sum_{\mu\nu}\sum_{q_j^{\mu\nu}}\bigg[q_j^{\mu\nu}\,\frac{n_{\nu}(q^{\mu\nu}_j)-n_{\mu}(q^{\mu\nu}_j)}{|\tilde{\xi}'_{\mu}(q^{\mu\nu}_j)-\tilde{\xi}'_{\nu}(q^{\mu\nu}_j)|}\nonumber\\
&&\hspace{10pt}\times\,\tilde{\mathcal{W}}_{\alpha\beta;\mu\nu}^{(\mathrm{d}+\mathrm{e})}(k^{\alpha\beta}_i,q^{\mu\nu}_j)\,\tilde{h}_{\mu\nu}(q^{\mu\nu}_j;\omega)\bigg],
\end{eqnarray}
where $\{k^{\alpha\beta}_i\}$ and $\{w^{\alpha\beta}_i\}$ are quadrature nodes and weights for the principal value integrals involving $\tilde{h}_{\alpha\beta;...}$, $N_p^{\alpha\beta}$ is the number of corresponding nodes and $\{q_j^{\mu\nu}\}$ is the set of solutions of $\Omega + \tilde{\xi}_{\nu}(q) - \tilde{\xi}_{\mu}(q)=0$.

In the case of finite swtiching rate ($\eta > 0$), $\omega = \Omega + i\eta$ and $\tilde{\Pi}^{(0)}_{\alpha\beta}(k,\omega)$ is regular on the real axis. Therefore, the last term in Eq.~(\ref{eq:quadh}) arising from the poles on the real axis will be absent.  

In our implementation, we imposed a large momentum cut-off on the integrals based on the same criteria as described in Sec.~\ref{sec:HFnum}. Quadratures for principal value integrations were generated using a combination of 16th-order Gauss-Lobatto quadrature in the proximity of the singular points and Simpson rule elsewhere~\cite{Abramowitz}. The resulting complex linear system of equations was solved for $h_{\alpha\beta}(k_i^{\alpha\beta},\omega)$ and consequently, the function $\mathcal{T}(\omega)$ was evaluated using the discretized version of Eq.~(\ref{eq:Tph}):
\begin{multline}\label{eq:Edothdisc}
\mathcal{T}(\omega) = -\eta^2\omega\sum_{\alpha\beta}\Bigg[\sum_{k_i^{\alpha\beta}}\frac{w_i^{\alpha\beta}k_i^{\alpha\beta}}{2\pi}\,\tilde{\mathbf{T}}_{\alpha\beta}(k_i^{\alpha\beta})\\
\times\,\frac{n_{\beta}(k^{\alpha\beta}_i)-n_{\alpha}(k^{\alpha\beta}_i)}{\omega + \tilde{\xi}_{\beta}(k^{\alpha\beta}_i) - \tilde{\xi}_{\alpha}(k^{\alpha\beta}_i)}\,\tilde{h}_{\alpha\beta;\gamma\lambda}(k^{\alpha\beta}_i;\omega)\\
-\frac{i}{2}\sum_{q_i^{\alpha\beta}}q_i^{\alpha\beta}\,\mathbf{T}_{\alpha\beta}(q_i^{\alpha\beta})\,\frac{n_{\beta}(q^{\alpha\beta}_i)-n_{\alpha}(q^{\alpha\beta}_i)}{|\tilde{\xi}'_{\alpha}(q^{\alpha\beta}_i)-\tilde{\xi}'_{\beta}(q^{\alpha\beta}_i)|}\nonumber\\
\times\,\tilde{h}_{\alpha\beta}(q_i^{\alpha\beta};\omega)\Bigg].
\end{multline}
Again, the last term is absent for finite switching rates.

Once $\tilde{h}_{\alpha\beta}(\omega)$ and $\mathcal{T}(\omega)$ are found, the enegry absorption spectrum, the spectral weight of excitons and the inter-subband transition rates can be readily evaluated using Eq.~(\ref{eq:Edot}),~(\ref{eq:colwght}) and~(\ref{eq:Iab}), respectively. For finite pulse switching rates, the excitons will be broadened and Eq.~(\ref{eq:Edot}) yields the full spectrum, including the broadened excitons and Eq.~(\ref{eq:colwght}) is no longer needed.

\newpage{\pagestyle{empty}\cleardoublepage}

\end{fmffile}

\end{document}